\documentclass[iop,twocolappendix,numberedappendix,appendixfloats]{emulateapj}

\usepackage{graphicx,threeparttablex,longtable,array} 
\usepackage{amssymb,amsmath,multirow,gensymb}
\usepackage{enumitem,multirow,rotating,placeins}
\usepackage{epstopdf}
\usepackage[caption=false]{subfig} 
\usepackage{natbib}
\newcommand{\mJybeam}{\mbox{mJy beam$^{-1}$}}
\newcommand{\uJybeam}{\mbox{$\mu$Jy beam$^{-1}$}}
\newcommand{\klambda}{\mbox{k$\lambda$}}
\newcommand{\vol}{\mbox{cm$^{-3}$}} 
\newcommand{\cden}{\mbox{cm$^{-2}$}} 
\newcommand{\kms}{\mbox{km s$^{-1}$}}
\newcommand{\um}{\mbox{$\mu$m}}
\newcommand{\ammonia}{\mbox{NH$_{3}$}}
\newcommand{\Msun}{\mbox{M$_{\odot}$}}

\newcommand{\cmg}{\mbox{cm$^2$ g$^{-1}$}}

\newcommand{\PI}{\mbox{$\mathcal{P}_I$}}
\newcommand{\sPI}{\mbox{$\sigma_{\mathcal{P}I}$}}
\newcommand{\PF}{\mbox{$\mathcal{P}_F$}}

\begin{document}

\title{Dust Polarization Toward Embedded Protostars in Ophiuchus with ALMA. II. IRAS 16293-2422}

\author{Sarah I. Sadavoy\altaffilmark{1$\dagger$}, 
Philip C. Myers\altaffilmark{1},
Ian W. Stephens\altaffilmark{1},
John Tobin\altaffilmark{2},
Woojin Kwon\altaffilmark{3,4},
Dominique Segura-Cox\altaffilmark{5},
Thomas Henning\altaffilmark{6},
Beno\^{i}t Commer\c{c}on\altaffilmark{7}
Leslie Looney\altaffilmark{8},
	 }

\footnotetext[$\dagger$]{Hubble Fellow}
	 
\altaffiltext{1}{Harvard-Smithsonian Center for Astrophysics, 60 Garden Street, Cambridge, MA, 02138, USA}
\altaffiltext{2}{National Radio Astronomy Observatory, 520 Edgemont Road, Charlottesville, VA 22903, USA}
\altaffiltext{3}{Korea Astronomy and Space Science Institute (KASI), 776 Daedeokdae-ro, Yuseong-gu, Daejeon 34055, Republic of Korea}
\altaffiltext{4}{Korea University of Science and Technology (UST), 217 Gajang-ro, Yuseong-gu, Daejeon 34113, Republic of Korea}
\altaffiltext{5}{Centre for Astrochemical Studies, Max-Planck-Institute for Extraterrestrial Physics, Giessenbachstrasse 1, 85748, Garching, Germany}
\altaffiltext{6}{Max-Planck-Institut f\"{u}r Astronomie (MPIA), K\"{o}nigstuhl 17, D-69117 Heidelberg, Germany}
\altaffiltext{7}{Universit\'{e} Lyon I, 46 All\'{e}e d'Italie, Ecole Normale Sup\'{e}rieure de Lyon, Lyon, Cedex 07, 69364, France}
\altaffiltext{8}{Department of Astronomy, University of Illinois, 1002 West Green Street, Urbana, IL, 61801, USA}


\date{Received ; accepted}

\begin{abstract}
We present high-resolution ($\sim 35$ au) ALMA Band 6 1.3 mm dust polarization observations of IRAS 16293.  These observations spatially resolve the dust polarization across the two protostellar sources and toward the filamentary structures between them.  The dust polarization and inferred magnetic field  have complicated structures throughout the region.  In particular, we find that the magnetic field is aligned parallel to three filamentary structures.  We characterize the physical properties of the filamentary structure that bridges IRAS 16293A and IRAS 16293B and estimate a magnetic field strength of 23-78 mG using the Davis-Chandrasekhar-Fermi method.  We construct a toy model for the bridge material assuming that the young stars dominate the mass and gravitational potential of the system.  We find that the expected gas flow to each star is of comparable order to the Alfv\'{e}n speed, which suggests that the field may be regulating the gas flow.  We also find that the bridging material should be depleted in $\sim 10^3$ yr.  If the bridge is part of the natal filament that formed the stars, then it must have accreted new material.  Alternatively, the bridge could be a transient structure.  Finally, we show that the 1.3 mm polarization morphology of the optically thick IRAS 16293B system is qualitatively similar to dust self-scattering.  Based on similar polarization measurements at 6.9 mm, we propose that IRAS 16293B has produced a substantial population of large dust grains with sizes between 200 and 2000 \um. 
\end{abstract}


\section{Introduction\label{Intro}}

Dust polarization is often used to trace magnetic fields in star-forming regions.  The polarization signatures are attributed to spinning, nonspherical dust grains that are preferentially aligned with their short axes parallel to the direction of the magnetic field \citep[e.g.,][]{DolginovMitrofanov76, HoangLazarian08}.  The cause of this alignment is still under debate, but in the most commonly accepted theory, the dust grains precess around the magnetic field lines due to radiative alignment torques (RATs) from an anisotropic radiation field  \citep[e.g.,][]{Andersson15}.  The resulting polarization structure is observed parallel to the magnetic field direction for stellar extinction and perpendicular to the magnetic field for thermal dust emission \citep{ChoLazarian07, Lazarian07}.   Dust is also ubiquitous in star-forming regions over multiple scales, meaning that observations of dust polarization can be used to possibly trace magnetic fields from clouds down to cores and protostellar disks.  

On cloud scales, previous observations from the extinction of background starlight  \citep[e.g.,][]{Goldsmith08, Palmeirim13, FrancoAlves15}  and thermal dust emission \citep[e.g.,][]{PlanckB15, Soler16, Soler17} give a bimodal relationship between the plane-of-sky magnetic field direction and the direction of parsec-sized filaments or elongated clouds based on density.  Low column density ($< 5 \times 10^{21}\ \cden$) clouds generally show inferred magnetic field orientations that are along their long axes, whereas high density ($> 5 \times 10^{21}\ \cden$) clouds have inferred magnetic fields that are orthogonal to their long axes.   This distinction in field orientation with cloud density is mainly attributed to how the gas condensed out of the interstellar medium.  The low-density clouds are non-self-gravitating and are produced by gas flowing along the field lines \citep[e.g.,][]{Hennebelle13, Klassen17, MoczBurkhart18}.  The high-density clouds, however, are self-gravitating and are formed by gravitational contraction that occurs preferentially along the field producing flattened structures with major axes perpendicular to the field lines \citep{MestelSpitzer56, Mouschovias76, NakamuraLi08, SolerHennebelle17}.   

On core scales, self-gravity is expected to dominate such that the flux-frozen magnetic field will be dragged inward, producing an hourglass structure \citep[e.g.,][]{MestelStrittmatter67, GalliShu93}.  If the magnetic field is weak, the contraction is mainly isotropic, resulting in a highly pinched hourglass morphology.   If the field is strong, collapse is suppressed in the direction perpendicular to the field, which limits the pinch and creates a flattened structure \citep{Field65, Crutcher12, Hull17}.  Previous observations have shown hourglass magnetic field morphologies toward both low-mass and high-mass cores indicative of a weak field \citep[e.g.,][]{Girart06, Rao09, Stephens13, Qiu14, Kandori17, Kwon18}. 

On disk scales, magnetic fields may be wrapped up into a toroidal morphology as the field lines are twisted by rotation in the disk, although a poloidal component may be necessary to drive disk winds and outflows \citep[e.g.,][]{BlandfordPayne82, McKeeOstriker07, Tomisaka11}.  Several recent studies, however, have shown that dust polarization in disks can also arise from other mechanisms independent of magnetic fields due to substantial grain growth.   Large ($> 10$ \um) dust grains can produce a polarized signature at (sub)millimeter wavelengths through Rayleigh self-scattering if the radiation field is anisotropic \citep[e.g.,][]{Kataoka15, Kataoka16, Pohl16, Yang16, Yang17}, or these grains can align with their long axes perpendicular to the flux gradient of radiation due to radiative torques \citep{LazarianHoang07, Tazaki17}.  The most recent high-resolution dust polarization observations of protostellar and protoplanetary disks from ALMA have shown polarization morphologies more indicative of these other mechanisms than magnetic fields \citep[e.g.,][]{Kataoka16hd, Kataoka17, Stephens17, Lee18, Cox18, Hull18, Harris18, Bacciotti18}.  Only a few sources have observations that are consistent with magnetically aligned dust grains on $< 100$ au scales;  HH 111 \citep{Lee18}, B335 \citep{Maury18}, VLA 1623 \citep{Sadavoy18}, BHB07-11 \citep{Alves18}, HD142527 \citep{Ohashi18}.  

Here we present high-resolution (35 au) 1.3 mm dust ALMA polarization observations of IRAS 16293-2422 (hereafter IRAS 16293) that show  highly ordered polarization vectors.  IRAS 16293 is a nearby \citep[$\approx$ 140 pc; ][]{OrtizLeon17, Dzib18}, very bright Class 0 protostellar system in the L1689 region of Ophiuchus.  It has two main sources (hereafter IRAS 16293A and IRAS 16293B) that are separated by $\sim 5$\arcsec\ or 700 au, although IRAS 16293A shows substructure at high resolution at radio frequencies \citep[e.g.,][]{Mundy92, Chandler05}.  IRAS 16293B is considered a single source with an optically thick, near face-on disk \citep{Pineda12,Zapata13,Oya18}. Both sources also host hot-core chemistry on $\lesssim 50$ au scales and have been the subject of numerous chemical studies \citep[e.g.,][]{Schoier02, Kuan04, Chandler05, Jorgensen11, Jorgensen16}.  

IRAS 16293 also has extended dust emission bridging IRAS 16293A and IRAS 16293B that was first identified in high resolution ALMA observations \citep[e.g.,][]{Pineda12, Jorgensen16}.   Several recent studies have attempted to model the bridge structure.  \citet{Jacobsen18} conducted a radiative transfer model assuming the bridge material was a curved cylinder that is heated on either end by the young stars.  van der Wiel et al. (submitted) expanded on this initial model using dense molecular gas tracers as an additional  constraint.  They proposed that the dust bridge is the remnant of the initial filamentary core from which the IRAS 16293A and IRAS 16293B stars formed.  

Here, we spatially resolve the dust polarization across this bridge structure for the first time.    We find uniform polarization vectors in this region that we attribute to magnetic grain alignment.  Rotating these polarization vectors by 90\degree, we find that the inferred plane-of-sky magnetic field direction of the bridge material is parallel to its filamentary shape.   Such parallel magnetic field orientations have not been previously seen at these scales.   This alignment was hinted at in previous, lower resolution observations \citep{Rao09, Rao14}, but is well resolved in the present ALMA observations.   In Section \ref{data}, we describe the ALMA observations.  In Section \ref{results} we show the continuum data, the observed  polarization structure, and the inferred magnetic field morphology.  In Section \ref{bfield_section}, we characterize the magnetic field orientation and strength in the bridge material.   In Section \ref{discussion}, we discuss the implications of the parallel field alignment  and the origins of the bridge material.  We also discuss the polarization of the IRAS 16293B disk in relation to dust self-scattering.  Finally, in Section \ref{summary}, we provide our conclusions.


\section{Observations}\label{data}

IRAS 16293 was observed in full polarization in Band 6 (1.3 mm, 233 GHz) with ALMA on 2017 May 20, July 11, and July 13 as part of a larger Cycle 3 (2015.1.01112.S) polarization survey of embedded protostellar systems in Ophiuchus.   Details of these observations are discussed in \citet{Sadavoy18}.  IRAS 16293 was observed with two separate pointings, with one field centered on IRAS 16293A and the other centered on IRAS 16293B.  Two pointings were used because IRAS 16293A and IRAS 16293B are $\approx 5$\arcsec\ apart \citep[e.g.,][]{Jorgensen16}, which is larger than the expected maximum recoverable scale for the array configuration (see below).  While we include two pointings to better recover the extended emission between the two stars, we caution that polarization observations of extended emission are reliable only within the inner third of the primary beam ($\sim 8\arcsec$).  The total time on each pointing was $\approx 7$ minutes.  Full details on the observations for the entire survey will be provided in a future article.

The two IRAS 16293 fields were cleaned and self-calibrated separately.  Following the same approach as outlined in \citet{Sadavoy18}, each field was cleaned interactively, and several self-calibration attempts were performed to test both phase and amplitude self calibration.  For the final maps, we first applied two rounds of phase-only self-calibration with decreasing solution intervals from infinity (full schedule block) to 30 seconds and then a single iteration of phase and amplitude self-calibration with a long (equivalent to infinite) solution interval.  Following all rounds of self-calibration, we examined the phase and amplitude solutions and the signal over all baselines to ensure that the solutions were good and the amplitudes were consistent from one iteration to the next.  For all iterations of \texttt{clean}, we use a robust weighting of 0.5, a taper of 0.1\arcsec, and the \texttt{multiscale} option with 1, 2.5, and 5 beams to better recover extended emission.   

For the final map, we performed a deep clean with both fields mosaicked together.  Since the emission detected toward IRAS 16293 is contained within the inner third of the primary beams for both fields, it is safe to mosaic these data (ALMA Helpdesk \#12418, private communication).   For the Stokes I observations, we ran \texttt{clean} in interactive mode with \texttt{multiscale} and user-defined masks, whereas for the Stokes Q, U, and V observations, we used noninteractive \texttt{clean}.  The final map sensitivities are 280 \uJybeam\ for Stokes I and 25 \uJybeam\ for Stokes Q, U, and V with an absolute flux calibration uncertainty of 10\%.  We note that the individual Stokes I, Q, and U maps from each field are identical to the the final mosaic observations with slightly better signal to noise.  Both the individual and mosaicked Stokes I maps are dynamic range limited because IRAS 16293 is so bright (peak S/N $\sim 2000$ in Stokes I), whereas the Stokes Q, U, and V maps have peak S/N $\lesssim 200$ and do not appear to be dynamic range limited.  The map resolution is $0.28\arcsec \times 0.23\arcsec$ ($\approx 35$ au) and the maximum recoverable scale is $2.6$\arcsec\ (360 au) as defined by a fifth percentile baseline of 102.275 m and an average elevation of 80\degree. 


\section{Results}\label{results}

\subsection{Continuum Results}

\begin{figure}[h!]
\includegraphics[width=0.475\textwidth,trim=1pt 1pt 1pt 1pt,clip=true]{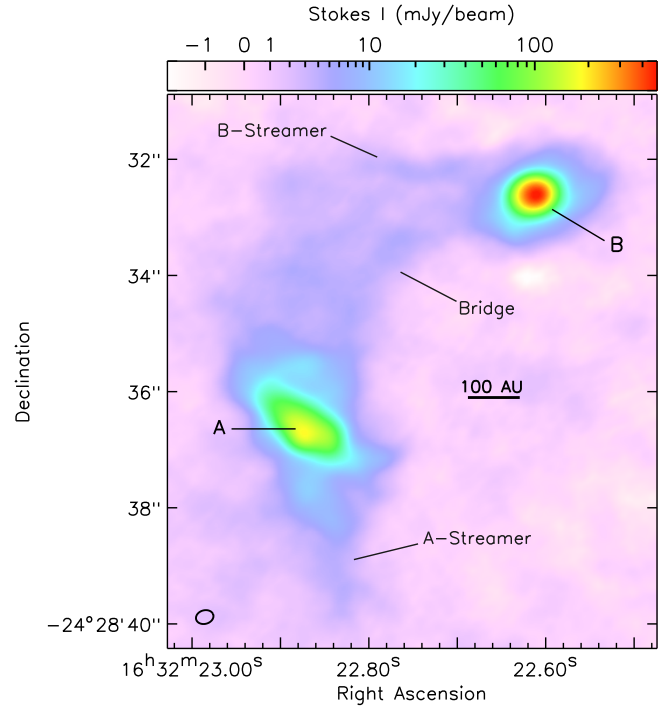}
\caption{ALMA 1.3 mm Stokes I observations of IRAS 16293.  The A and B sources are labeled.  The synthesized beam size is in the lower-left corner.  \label{stokesI}}
\end{figure}

Figure \ref{stokesI} shows the Stokes I mosaic map of IRAS 16293 with the main sources, A and B, labeled.  The figure also shows extended continuum emission between the two sources.  This dust emission has been seen previously from ALMA observations, with the material between the two sources identified as a ``bridge'' and the extensions from IRAS 16293A and IRAS 16293B as ``streamers'' \citep[e.g.,][]{Pineda12, Jorgensen16}.  We use this same terminology here and refer to the material connecting the two sources as the Bridge, the extension south from IRAS 16293A as the A-Streamer, the extension east from IRAS 16293B as the B-Streamer.   Figure \ref{stokesI} further resolves the Bridge into two parallel structures that are also seen in \citet{Jorgensen16}.  

Figure \ref{stokesI} shows very distinct morphologies for IRAS 16293A and IRAS 16293B.  IRAS 16293A is elongated and highly extended, whereas IRAS 16293B is compact and circular.  These morphologies agree well with previous ALMA observations of the dust continuum for this region \citep[e.g.,][]{Pineda12, Jorgensen16, Jacobsen18}.  In general, IRAS 16293A is thought to be an edge-on source and IRAS 16293B is consistent with  a face-on disk geometry \citep{Oya18}.  

\begin{figure}[h!]
\includegraphics[width=0.475\textwidth,trim=1pt 1pt 1pt 1pt,clip=true]{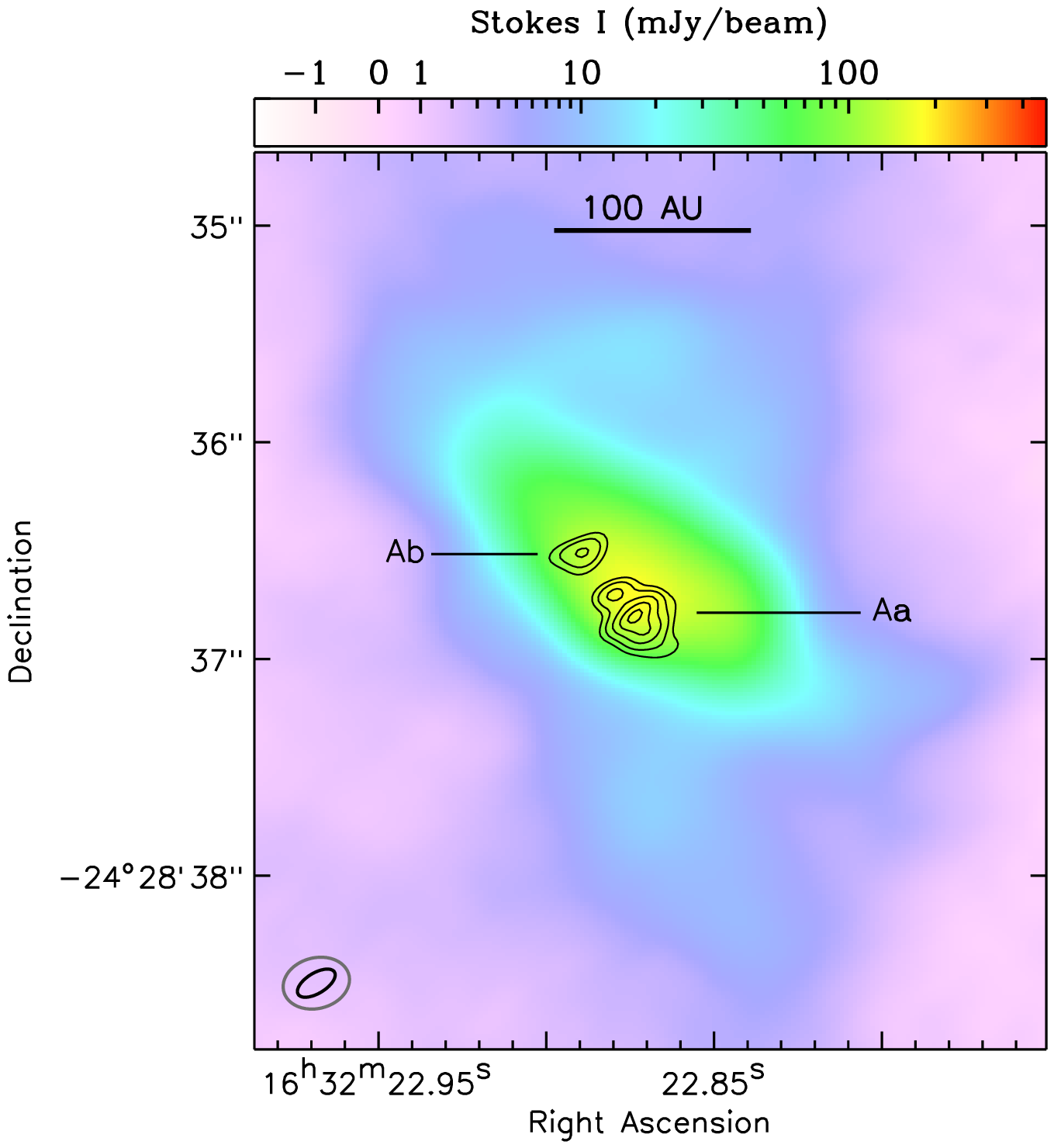}
\caption{High-resolution image of IRAS 16293A (contours) overlaid on the full Stokes I continuum data.  The higher resolution map was created using $uv$ distances $> 600$ \klambda\ and uniform weighting to remove the extended emission around the two sources.  The contours show levels of 25\%, 40\%, 60\%, 80\%, and 95\%, of the peak intensity.  The Aa and Ab components \citep{Chandler05, Chen13} are resolved in the filtered map.  The synthesized beam sizes for the background image (grey) and the contours (black) are given in the lower-left corner.  \label{highres}}
\end{figure}

The continuum map of IRAS 16293A is relatively smooth.  Previous observations, however, found multiple substructures in observations at 15-300 GHz and resolutions of $\approx 0.4$\arcsec\ \citep[e.g.,][]{Wootten89, Chandler05, Loinard13, Chen13}.   Two substructures in particular, Aa and Ab, appear to trace thermal dust emission indicative of two distinct stars   These sources are not seen in Figure \ref{stokesI} due to confusion with bright, extended emission at shorter baselines.   We remove this extended emission using \texttt{clean} with $uv$ distances $ > 600$ \klambda\ and uniform weighting. Figure \ref{highres} shows contours from this higher resolution map (beam of 0.18\arcsec\ $\times$ 0.09\arcsec; 25 au $\times$ 13 au) on the full Stokes I image of IRAS 16293A (e.g., a zoom-in from Figure \ref{stokesI}).  Indeed, by imaging only the largest baselines, we resolve the Aa and Ab continuum structures from previous interferometric studies \citep{Chandler05, Chen13}.    

Table \ref{positions} gives the positions of IRAS 16293A and IRAS 16293B from fitting Gaussian functions to each object in CASA.  We also include the positions of each component from the higher resolution map shown in Figure \ref{highres}.  We note that a second peak is seen between Aa and Ab, which could be associated with the A1 or A2$\beta$ radio sources \citep{Loinard13}.  Identification of this structure is unclear, however, given that no other (sub)millimeter study found a corresponding detection toward these radio sources and that the radio emission in the region has complicated motions from different ejecta \citep[e.g.,][]{Imai07,Pech10}. 

{\setlength{\extrarowheight}{1pt}%
\begin{table}[h!]
\caption{Position of compact sources in IRAS 16293}\label{positions}
\begin{tabular}{llll}
\hline\hline
				&	R.A. (J2000)		& Decl. (J2000)	 & Beam (arcsec) \\ 
\hline
\multicolumn{4}{c}{Full Map}\\
\hline
IRAS 16293A		&	16:32:22.873		&  -24:28:36.63	  & $0.28\times 0.23$ \\ 
IRAS 16293B		& 	16:32:22.613		&  -24:28:32.61  & $0.28\times 0.23$ \\ 
\hline
\multicolumn{4}{c}{High-Resolution Map} \\
\hline
IRAS	 16293Aa	\tablenotemark{$\star$}	&	16:32:22.880		& -24:28:36.71  & $0.18 \times 0.09$ 	\\
				&	16:32:22.874		& -24:28:36.81  & $0.18 \times 0.09$ 	\\
IRAS 16293Ab		&	16:32:22.890		& -24:28:36.52	& $0.18 \times 0.09$\\
IRAS 16293B		&	16:32:22.607		& -24:28:32.64	& $0.18 \times 0.09$\\
\hline
\end{tabular}
\begin{tablenotes}[normal,flushleft]
\item \tablenotemark{$\star$} {We give two positions for IRAS 16293Aa centered on the two peaks seen in Figure \ref{highres}.}
\end{tablenotes}
\end{table}
}

\subsection{Polarization Results}

The Stokes Q and U observations are combined to calculate the polarization intensity (\PI), position angles ($\theta$), and fraction ($\PF$).  The polarization intensity is determined from $\PI = \sqrt{Q^2 + U^2}$.  In regions where the Stokes Q and U data are not well detected, the polarized intensity has a positive bias that must be removed.  We use a basic maximum likelihood characterization to debias the polarized intensities \citep[e.g., following,][]{SimmonsStewart85, Vaillancourt06},

\begin{equation}
\PI = \sqrt{Q^2+U^2 - \sPI^2} \label{debiasEq},
\end{equation}
where $Q$ and $U$ are the Stokes Q and U intensities at each pixel and $\sPI$ is the noise in the polarization map.  We assume $\sPI \approx \sigma_{Q} \approx \sigma_U$ for simplicity.   This approach to debias dust polarization is reliable for well-detected measurements with $\PI/\sPI >4$  \citep{Vaillancourt06}. 

The polarization position angle and polarization fraction are defined as
\begin{equation}
\theta = \frac{1}{2}\tan^{-1}\frac{U}{Q}  \label{ang_eq}
\end{equation}
\begin{equation} 
\PF = \frac{\PI}{I}.
\end{equation}
The polarization position angles are measured from $-90\degree$ to 90\degree, North to East.  We also assume an error of
\begin{equation}
\sigma_{\theta} = \frac{1}{2}\frac{\sPI}{\PI} \label{ang_eq_err}
\end{equation}
for the position angles \citep{Hull14}, which corresponds to $\sigma_{\theta} \lesssim 7$\degree\ for $\PI/\sPI >4$.  We adopt a polarization fraction uncertainty of 0.1\%\ for the instrument polarization.  This polarization fraction error is considered appropriate for extended emission within the inner third of the primary beam\footnote{ALMA Technical Handbook.}.  

\begin{figure*}[t!]
\centering
\includegraphics[width=0.98\textwidth,trim=1pt 1.8cm 1pt 1pt,clip=true]{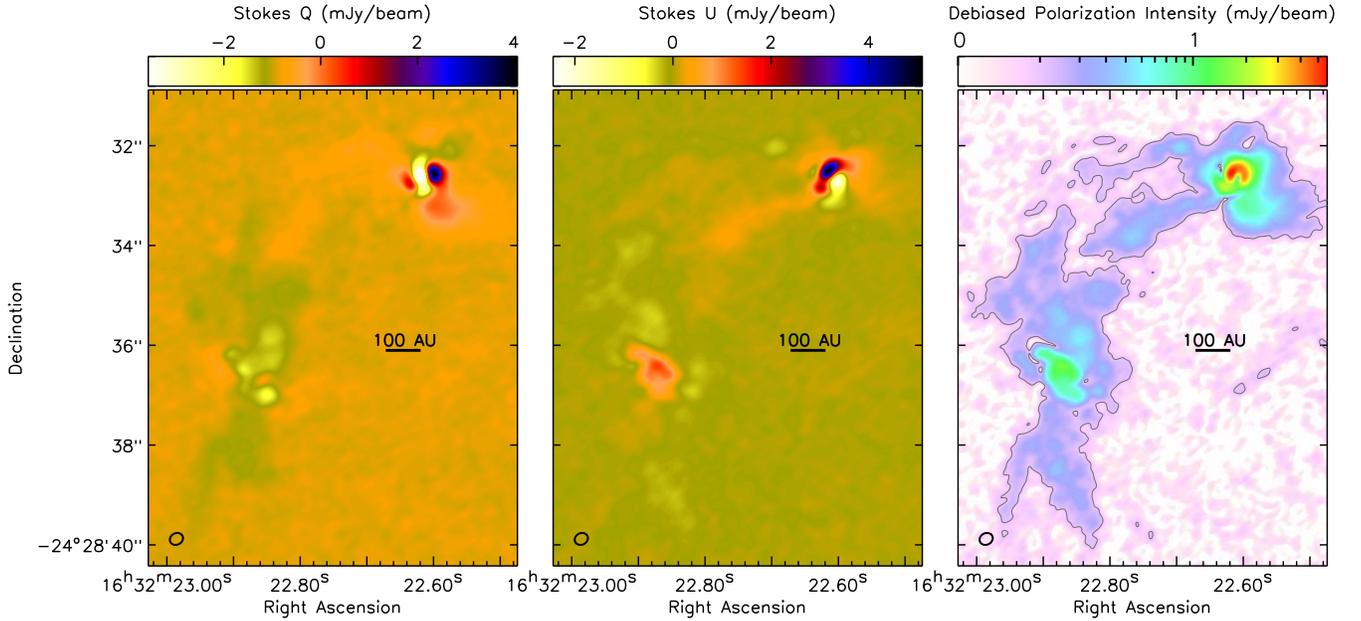} 
\caption{Maps of Stokes Q (left), Stokes U (middle) and debiased polarized intensity (right) for IRAS 16293.  The thin contours in the right panel show a polarized intensity level of $4\ \sPI$.  Note that the polarized intensity map uses a log scaling, whereas the Stokes Q and U maps use linear scaling.  The synthesized beam size is in the lower-left corner.  \label{polarization}}
\end{figure*}

Figure \ref{polarization} shows the Stokes Q, Stokes U, and the debaised polarization intensity maps of IRAS 16293.   We see substantial structure in the individual Stokes Q and U maps for both protostars.  We note that the two individual fields for IRAS 16293 gave consistent results for Stokes I, Q, and U.  We also find strong detections $> 30\sigma$ of Stokes V toward IRAS 16293B, but the two individual fields gave inconsistent values.  The emission changes from positive in one field to negative in the other field with different magnitudes.  At the time of these observations, Stokes V data were not well constrained by ALMA.  As such, we do not use the Stokes V data in this study.

\begin{figure*}[t!]
\includegraphics[height=10.5cm,trim=1pt 1pt 1pt 1pt,clip=true]{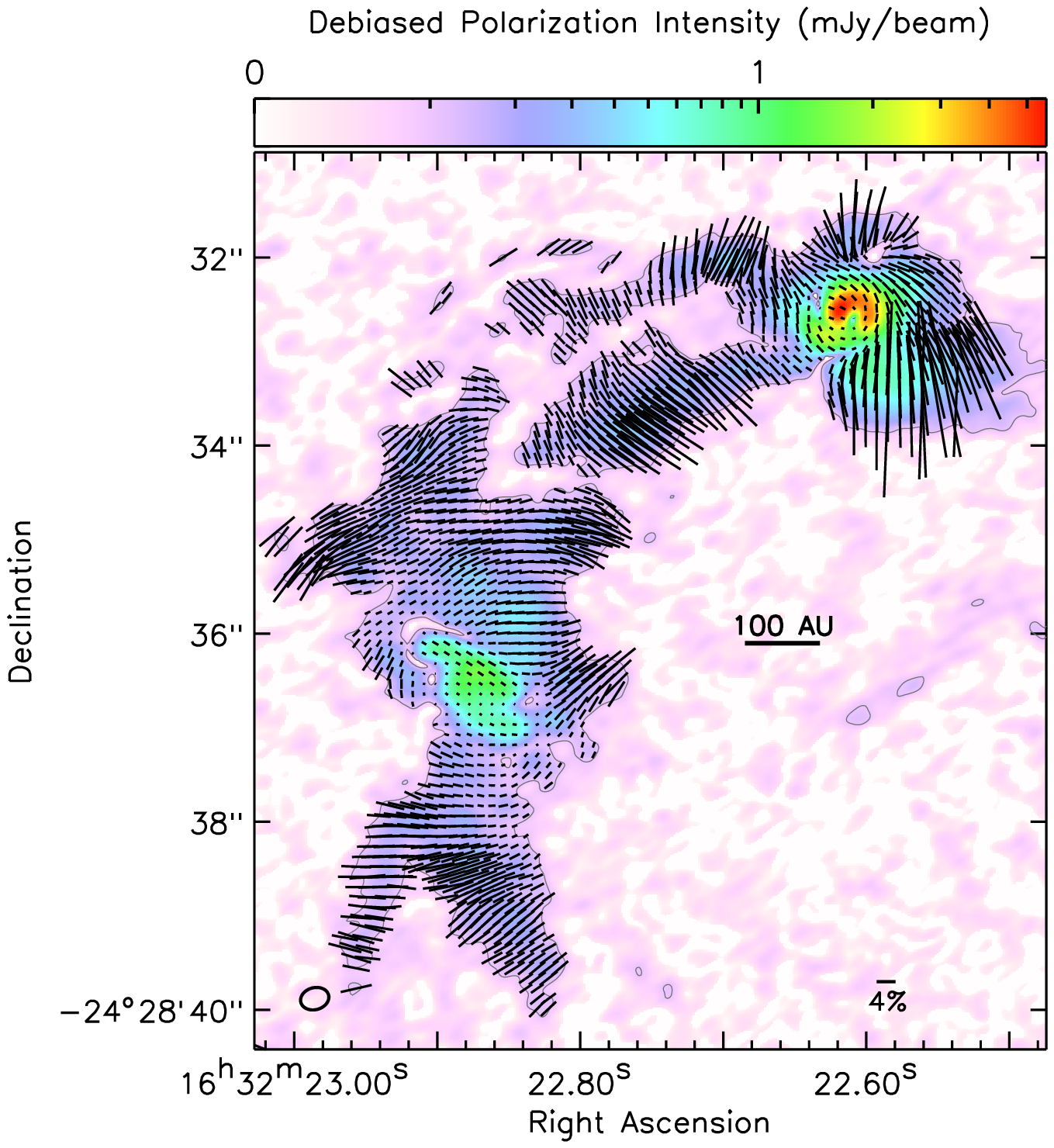}\quad
\includegraphics[height=10.5cm,trim=1pt 1pt 1pt 1pt,clip=true]{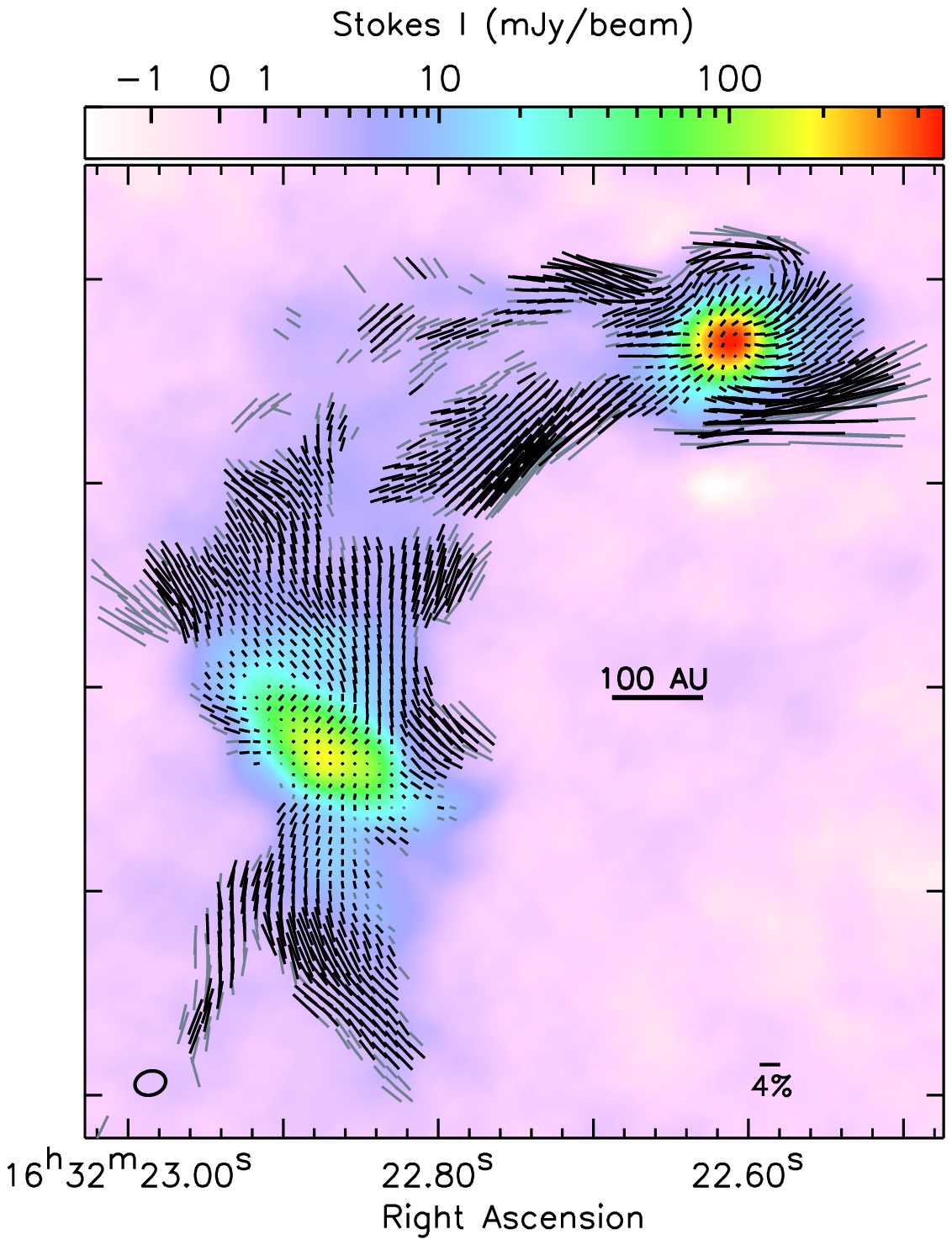}
\caption{\emph{Left:}  Polarization e-vectors overlaid on the debiased polarization intensity map.  All vectors correspond to $I > 3\ \sigma_I$ and $\PI > 4\ \sPI$ and are Nyquist sampled along the major axis.   \emph{Right:} Inferred magnetic field direction of IRAS 16293 overlaid on the Stokes I intensity map.    The same polarization e-vectors from the left panel are rotated by by 90\degree\ to show the b-vector morphology.  The vectors in black represent the most robust b-vectors with $I > 5\ \sigma_I$ and $\PI > 5\ \sPI$.  For both panels, the vector lengths are scaled by their polarization fraction, with a reference scale of $\PF = 4$\%\ given in the lower-right corner.  The synthesized beam size is in the lower-left corner.  \label{bfield}}
\end{figure*}

Figure \ref{bfield} shows the polarization morphology and the inferred magnetic field orientation for IRAS 16293 from our ALMA observations.  Hereafter, we use the term ``e-vector'' to refer to the unrotated, observed polarization position angles and ``b-vector'' to refer to the rotated, inferred magnetic field orientation.  We note that these are not true vectors, as dust polarization only gives a polarization angle with an 180 degree ambiguity.    We show the e-vectors and b-vectors for regions with $I > 3\ \sigma_I$ and $\PI > 4\ \sPI$, where the black b-vectors  correspond to  regions with $I > 5\ \sigma_I$ and $\PI > 5\ \sPI$.   Both the e-vectors and b-vectors have been scaled by their corresponding polarization fraction, with a reference of 4\%\ given in the lower-right corner.    

\subsection{Inferred Magnetic Field Morphology}\label{bfield_morphology}

Figure \ref{bfield} shows the most detailed polarization map of IRAS 16293 to date.  The polarization and inferred magnetic field morphologies are both highly organized.  The global pattern matches well what has been seen at coarser resolutions with the Submillimeter Array \citep[SMA;][]{Rao09, Rao14, Galametz18}, but the higher resolution data presented here show distinctly uniform structures toward IRAS 16293A and IRAS 16293B as well as the Bridge and two streamers (e.g., see Section \ref{filfinder}).   There is also a zone of distinctly unpolarized dust at a 4 $\sigma$ level through the Bridge and between the Bridge and the B-Streamer.    This depolarization zone spans roughly one beam and passes through the Bridge near the center point between the two stars.  We note that the dust emission remains depolarized even if we relax our criteria to $\sPI/\PI > 3$.  We discuss the depolarization in Section \ref{depol}. 

The magnetic field morphology is highly organized with distinct structures seen around each source and a parallel morphology along the filamentary Bridge, A-Streamer, and B-Streamer.   There are hints of the hourglass magnetic field shape that was previously identified toward IRAS 16293A by \citet{Rao09} using polarization data from the SMA.   With our higher resolution observations, however, this pinched pattern breaks up into a more complicated morphology.  The polarization directly on the brightest emission of IRAS 16293A is relatively uniform with b-vector position angles of $\sim -40$\degree.  This position angle changes significantly away from this main part.  For the A-Streamer, the position angles are mainly between $\sim 0-40$\degree, whereas above IRAS 16293A, the polarization smoothly changes from 40\degree\ to 0\degree\ to 60\degree\ going from East to West.  We note that some of this magnetic field structure is evident in \citet{Rao09}.  Thus, hourglass shapes detected on coarse resolutions may not truly trace smoothly contracting fields when examined at higher resolution.  

IRAS 16293B also shows hints of pinched magnetic field lines at its periphery.  This pinched structure is seen outside of the the compact, optically thick, dense hot core and disk.  The field shows a distinct inward curvature that is separate from B-Streamer.  Nevertheless, we caution that the magnetic field structure toward IRAS 16293B is not well characterized by our observations.  First, we lack observations on shorter baselines to trace fully the magnetic field morphology at large angular extents.  The pinched pattern is only hinted at with a few vectors and may be confused by the B-Streamer.   Second, the magnetic field morphology toward the bright IRAS 16293B hot core and disk is likely complicated by polarization signatures from dust self-scattering.  We discuss polarized self-scattering in Section \ref{polB}.


\section{Magnetic Field in the Filamentary Structures} \label{bfield_section}

One of the most striking features in Figure \ref{bfield} is the alignment between the field orientation and the filamentary structures in IRAS 16293.  Figure \ref{bfield} shows that the projected plane-of-sky magnetic field is parallel to the Bridge, the A-Streamer, and the B-Streamer.   In this Section, we characterize these filamentary structures using a filament-finding algorithm and then compare their orientation to the position angles of the neighboring b-vectors.

\subsection{Filamentary Structure Properties}\label{filfinder}

We use the filament-finding algorithm FilFinder2D \citep{KochRosolowsky15} to produce a skeleton diagram of all filamentary structures in IRAS 16293.  Since  FilFinder2D is designed to identify large-scale filaments with widths of $\sim$ 0.1 pc rather than small-scale structures with widths of $< 100$ au (E. Koch 2018, private communication), we had to construct a user-defined mask to allow the program to work on 100 au scales.  The user-defined mask is constructed using a mask of $I > 5\ \sigma_I$ to select the brightest emission associated with IRAS 16293.  We further apply an intensity cut of $I < 9\ \sigma_I$ in the small region between the Bridge and B-Streamer to account for the gap between them, and also an intensity cut of $I < 12\ \sigma_I$ near IRAS 16293A and IRAS 16293B to remove bright emission around the protostars that biases the algorithm.  These thresholds are chosen so that the resulting skeletons by eye trace the curvature of the Bridge and streamers.      

\begin{figure}[h!]
\includegraphics[width=0.475\textwidth,trim=1pt 1pt 1pt 1pt,clip=true]{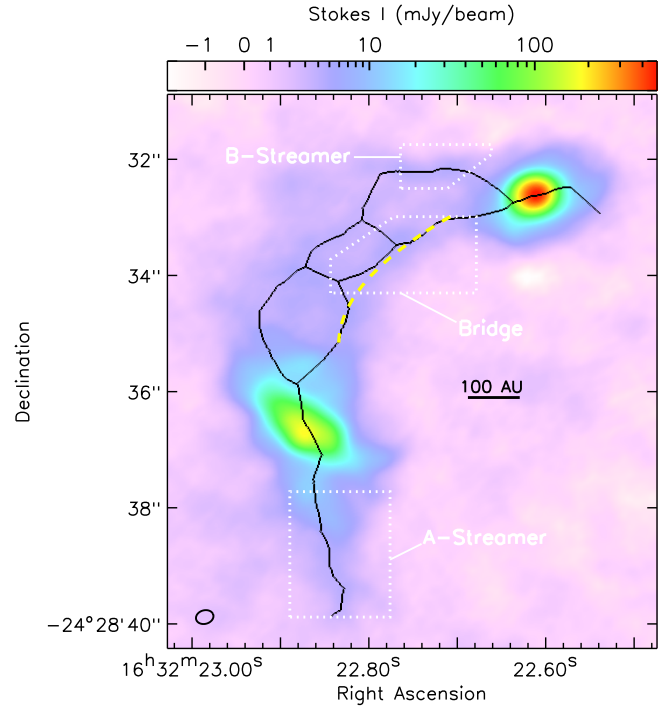}
\caption{Skeleton outlines for the filamentary structures in IRAS 16293 from FilFinder2D.  The background image shows the Stokes I map (see Figure \ref{stokesI}), and the black lines show the skeleton outlines.  The three regions further discussed in the text are labeled with dotted boxes to show the regions used in Section \ref{parallel}.  The dashed yellow curve shows the best-fit function for the main Bridge skeleton, which was too jagged to use directly in Section \ref{parallel} (see text).  The synthesized beam size is in the lower-left corner.  \label{skeleton}}
\end{figure}

Figure \ref{skeleton} shows the skeleton outline of IRAS 16293 from running FilFinder2D on our data with our user-defined mask.  By eye, the skeleton outline has good agreement with the Stokes I emission.  We clearly recover the Bridge, A-Streamer, and B-Streamer as shown in Figure \ref{stokesI} and \citet{Jorgensen16}.  The Bridge material breaks up into multiple filamentary structures with two main parallel branches (the northern one connects to the B-Streamer) and several connecting subbranches.  Hereafter, we focus on the southern Bridge branch (see Figure \ref{skeleton}), which is the brightest and most polarized branch (see Figure \ref{bfield}).   The southern Bridge skeleton is a bit jagged because of the subbranches, however.  To represent its continuous shape with a smooth function, we fit a simple $y = \sqrt{x}$ curve to the skeleton.  The dashed yellow curve in Figure \ref{skeleton} shows this best-fit function. 

\begin{figure}[h!]
\includegraphics[width=0.475\textwidth,trim=1pt 1pt 1pt 1pt,clip=true]{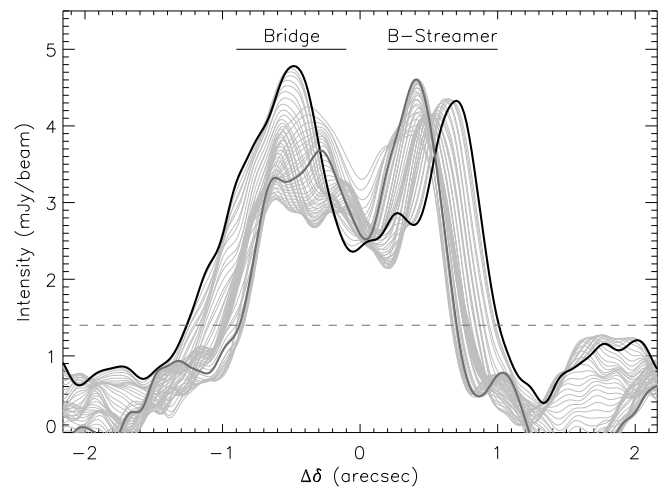}
\caption{Stokes I intensity slices in the vertical direction through the Bridge and B-Streamer.  The reference position for the slice is the midpoint between the two skeletons.  The solid black curve shows the slice through the peak of the Bridge and the dark grey curve shows the slice through the peak of the B-Streamer.  The dashed horizontal line shows a 5$\sigma_I$ value. \label{slice}}
\end{figure}

Although the main Bridge branches have similar intensities, there are clear, resolved dips in intensity between them.  Figure \ref{slice} shows vertical intensity slices through the Bridge and B-Streamer over a range across the area of the B-Streamer as defined in Figure \ref{skeleton}.  The reference point for all the slices is the midpoint between the Bridge and B-Streamer skeletons (i.e., where the dips reach a minimum intensity).  Overall, we find that the intensity drops by up to a factor of two between the Bridge and B-Streamer.  

\subsection{Field Orientation}\label{parallel}

 We quantify the degree of agreement between the magnetic field orientation and the filamentary dust structures in IRAS 16293 by measuring the difference in angle between the dust structure skeleton and the b-vectors.  For simplicity, we consider only those pixels within the regions shown in Figure \ref{skeleton} within the dotted boxes.  The dotted boxes correspond represent regions with sufficient b-vectors that have $\PI/\sPI > 5$ and $I/\sigma_I > 5$ (see Figure \ref{bfield}). 

\begin{figure*}[t!]
\includegraphics[width=0.32\textwidth,trim=1pt 1pt 1pt 1pt,clip=true]{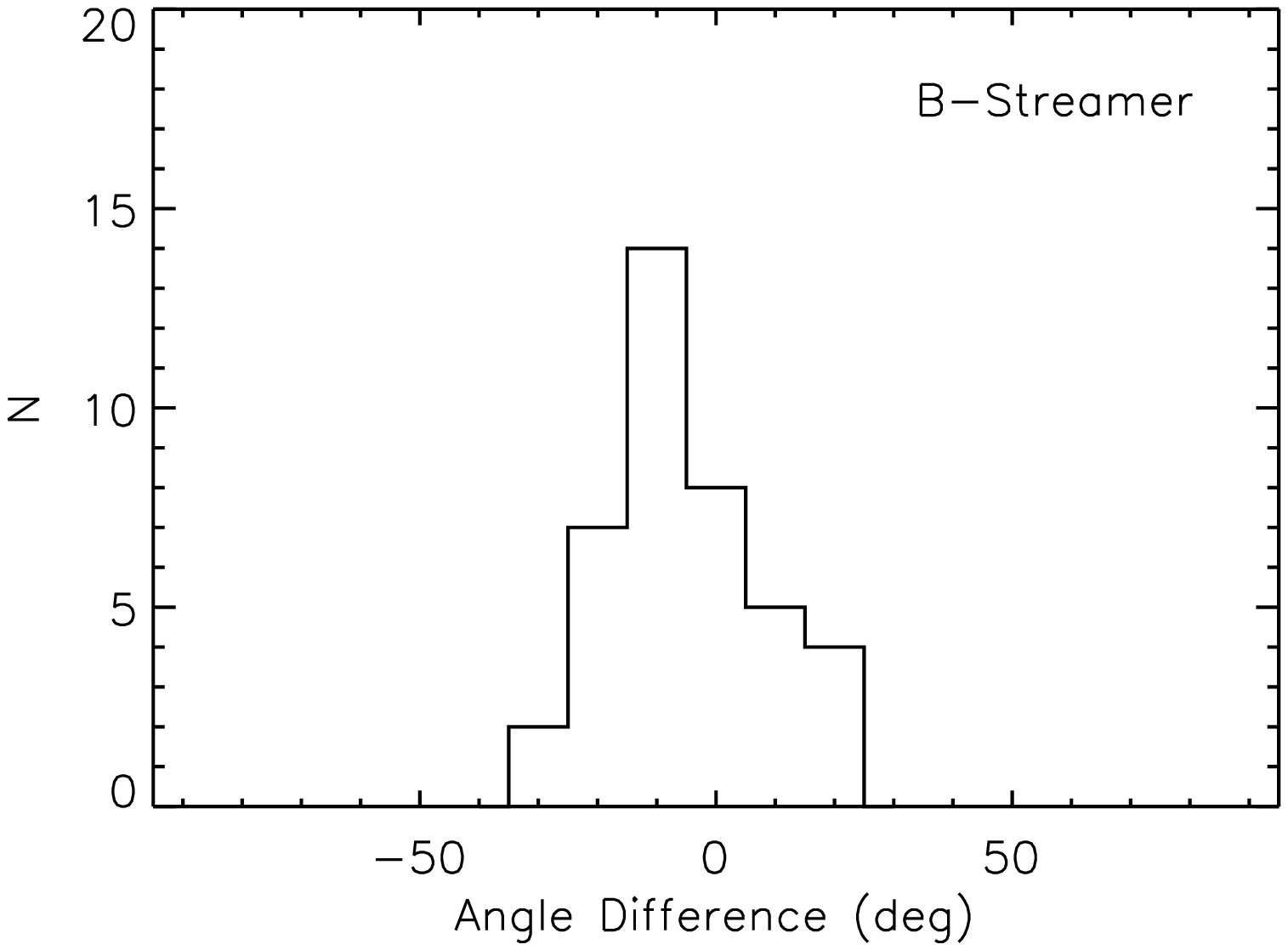}
\includegraphics[width=0.32\textwidth,trim=1pt 1pt 1pt 1pt,clip=true]{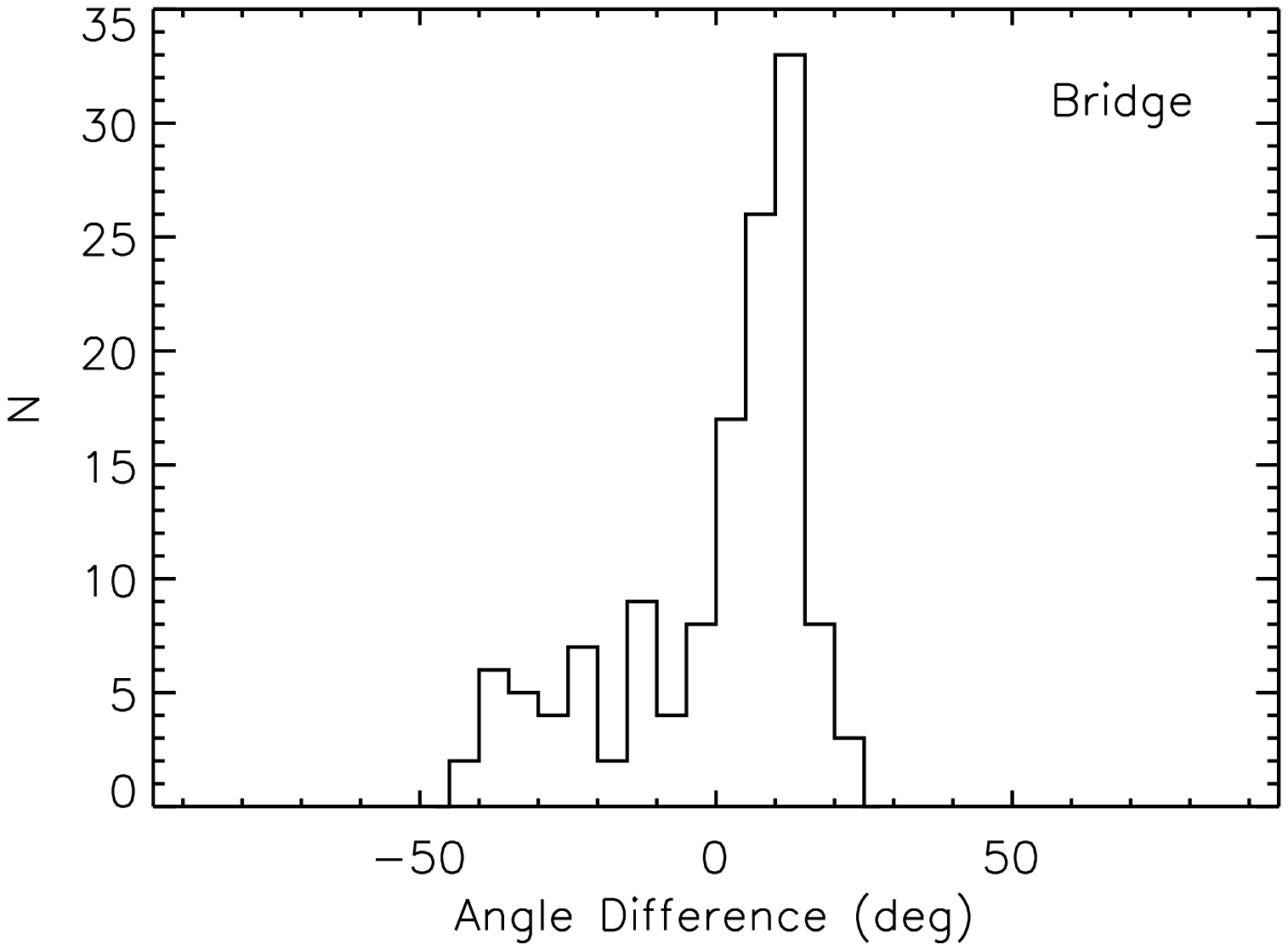}
\includegraphics[width=0.32\textwidth,trim=1pt 1pt 1pt 1pt,clip=true]{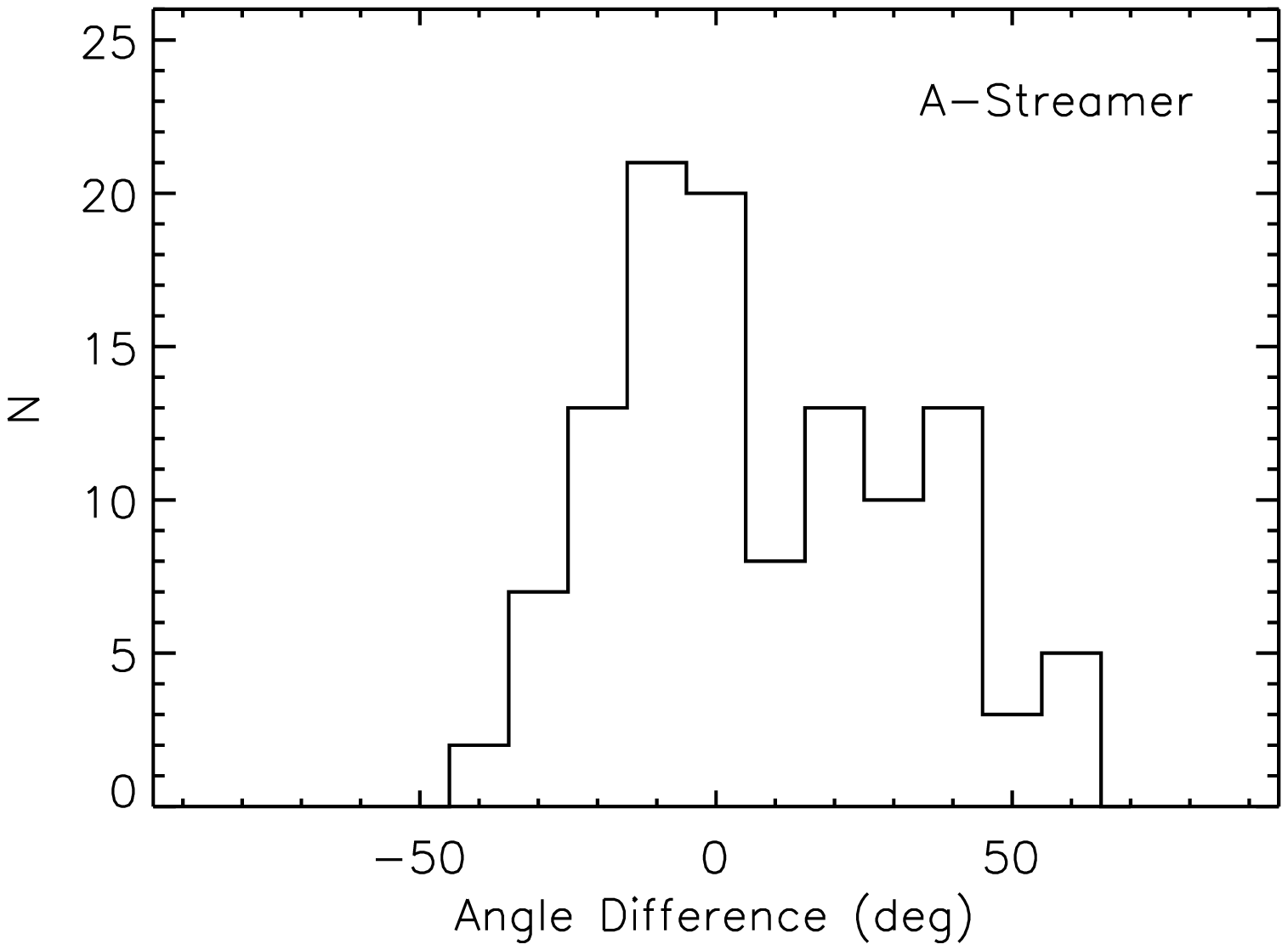}
\caption{The alignment between the filamentary structures and magnetic field orientation.  An angle of 0\degree\ corresponds to parallel and 90\degree\ corresponds to perpendicular.  The three panels show the B-Streamer, southern Bridge region, and the A-streamer, respectively.  The thick histograms show the angle difference between the observed b-vector position angles and the direction of each elongated structure within the regions outlined in Figure \ref{skeleton}. \label{angles}}
\end{figure*}

Figure \ref{angles} shows the distribution of angle difference between the skeleton slope and b-vector orientation for the B-streamer (left), Bridge (middle), and A-Streamer (right).  For each pixel in the defined regions, we identify the closest skeleton and measure the slope at that position.  For the Bridge, we use the derivative of the dashed curve in Figure \ref{skeleton} to measure its slope, whereas for the A-streamer and B-streamer, we use the slope of the skeleton from a linear-least squares fit to 5 pixels centered on the nearest pixel.  The angle differences are defined as 0\degree\ corresponding to parallel and 90\degree\ corresponding to perpendicular.  Figure \ref{angles} shows that the magnetic field morphology is primarily oriented within 30\degree\ of the filamentary structures in IRAS 16293 and that the distributions peak near 0\degree.  The Bridge and B-Streamer show the most peaked profiles, whereas the A-Streamer has a broader distribution.  

\begin{figure}
\includegraphics[width=0.475\textwidth,trim=1pt 1pt 1pt 1pt,clip=true]{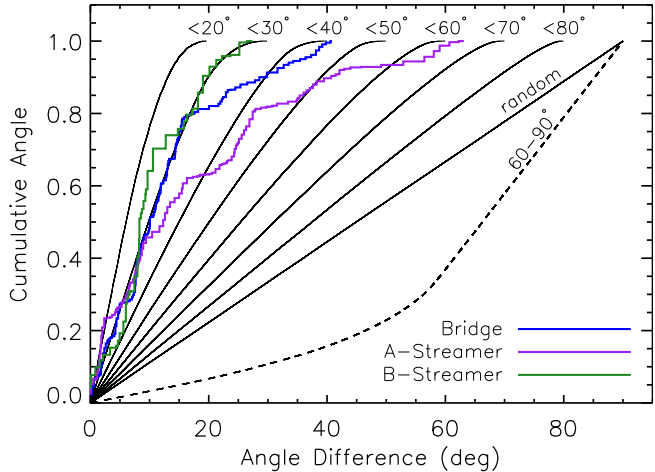}
\caption{The cumulative distribution of angle difference in each region.  The blue, purple, and green curves show the profiles for the Bridge, A-Streamer, and B-Streamer, respectively.  The black curves show the projected 2-D angle differences relative to a fixed axis from a Monte Carlo sample of 100,000 randomly selected 3-D vectors.  The solid curve shows the expected profiles for a subsample of vectors with 3-D orientations that range from $0-20$\degree\ to $0-90$\degree\ in steps of 10\degree\ for the upper bound.  The profile with vectors between 0 and 90\degree\ of the fixed axis represents a random sample and is labeled as such.  The dashed curve shows the equivalent distribution for a mainly perpendicular magnetic field with vectors between $60-90$\degree\ only.    \label{cum_angles}}
\end{figure}

Figure \ref{cum_angles} compares the cumulative angle difference profiles (colored curves) of each region with the expected profiles for a random, parallel, and perpendicular magnetic field (black curves) using a large Monte Carlo sample.  This figure is a variation of the Monte Carlo samples used in \citet{Hull14} and \citet{Stephens17outflow}.  First, we generate a large Monte Carlo sample of 100,000 randomly defined 3-D vectors to represent the field orientation.   Second, we project these vectors onto 2-D to determine the plane-of-sky field orientation.  Third, we measure the angle difference between the projected magnetic field vector and a fixed vector that represents the filamentary structures.  For simplicity, we assume that the filamentary structures are all in the plane-of-the-sky and can be represented by a single position angle on the sky.  Figure \ref{cum_angles} shows nine cases from this Monte Carlo sample.  The eight solid lines represent the expected profiles for subsamples of vectors with 3-D orientations that range from $0-20\degree$\  to $0-90$\degree\ in steps of 10\degree\ for the upper bound.  The case where the vectors are between $0-90$\degree\ represents a random magnetic field orientation.  The dashed curve shows the profile for a perpendicular magnetic field, where the 3-D vectors are between $60-90$\degree\ from the fixed axis.  The Bridge and B-Streamer have profiles that are most similar to $0-30$\degree\ distribution and the A-streamer mostly has vectors between $0-50$\degree. We note that the observed profiles in Figure \ref{cum_angles} are measured from Nyquist sampling the magnetic field position angles within the beam and as such, they do not represent completely independent samples.  Nevertheless, even if we undersample the observations by taking only one data point per beam, we still find profiles that parallel distributions with angles $< 30-50$\degree\ and that all three distributions deviate significantly from both random and perpendicular.   

Figures \ref{angles} and \ref{cum_angles} show that the magnetic fields in the filamentary structures are generally biased toward a parallel morphology.  To quantitatively test these distributions, we conduct the two-sample Kolmogorov-Smirnov (KS) test \citep{Kolmogorov33,Smirnov48,Massey51} and the Anderson-Darling (AD) test \citep{AndersonDarling52, AndersonDarling54} between the observed distributions and the projected 3-D distributions from the Monte Carlo sample (see Figure \ref{cum_angles}).  These statistical tests assume a null hypothesis that the observed and modeled data are drawn from the same sample.  This null hypothesis is rejected if the distribution functions of the two samples have a large deviation.  While generally similar in practice, the AD test is more sensitive to the tails of the distribution function than the KS test, making it a more powerful statistic \citep{Stephens74, Hou09}.  We use the KS test as a more popular measure and the AD test to ensure that we can robustly reject or accept the null hypothesis for each of the assumed distributions.

For the perpendicular and random cases, we find very similar, low $p$-values\footnote{Low $p$-values indicate that you can significantly reject the null hypothesis.} of $\ll 10^{-10}$ from both the AD and KS tests.   Thus, the magnetic field is statistically inconsistent with randomly aligned or primarily perpendicular morphologies in each of the filamentary structures.  For the parallel cases, we find that the B-Streamer is statistically consistent with a distribution of $<30\degree$ ($p \approx 0.4-0.5$ from the AD and KS tests) and we cannot reject the null hypothesis.  The KS test gives a marginally critical value between the A-Streamer and the distribution of $<50\degree$ ($p \approx 0.06$) at the 5\%\ level, but the AD test indicates these two distributions are statistically inconsistent.  Since the A-Streamer distribution has a long tail (see Figure \ref{angles}), we use the AD test results instead and reject the null hypothesis.    Both the KS test and the AD test indicate that the Bridge is inconsistent ($p < 0.005$) with all of the tested distributions, which is not surprising given the tail in its angle histogram (see Figure \ref{angles}).  Thus, the magnetic field morphologies of the filamentary structures are mainly parallel, but the field orientations in the A-Streamer and Bridge are statistically more complex than a simple, mostly parallel distribution.   

\subsection{Field Strength}\label{strength}

We estimate a magnetic field strength for the Bridge using the Davis-Chandrasekhar-Fermi (DCF) method \citep{Davis51, ChandrasekharFermi53}.  In brief, this method assumes that any fluctuations in the magnetic field is produced by turbulence in the cloud.  These fluctuations can be estimated by the dispersion in polarization angle, where higher degrees of dispersion indicate a more random magnetic field.    Following \cite{Crutcher04}, we calculate the plane-of-sky magnetic field strength as,
\begin{equation}
B = Q \sqrt{4\pi\rho}\frac{\sigma_v}{\sigma_{\theta}},
\end{equation}
where $\rho$ is the gas density, $\sigma_v$ is the velocity dispersion of the gas, $\sigma_{\theta}$ is the dispersion in polarization angle, and $Q$ is a scaling factor to account for the three-dimensional project effects.  We adopt $Q = 0.5$ \citep{Ostriker01}.  This equation can be written as,
\begin{equation}
B = 22 \mbox{mG} \left(\frac{n}{10^6\ \vol}\right)^{1/2}\left(\frac{\sigma_v}{1\ \kms}\right)\left(\frac{\sigma_{\theta}}{\mbox{deg}}\right)^{-1}.
\end{equation}

We focus on the southern Bridge branch, because it is the longest, brightest, and most coherent filamentary structure.  For the velocity dispersion, we adopt $\sigma_v = 0.75$ \kms\ from C$^{17}$O (3-2) line emission (M. van der Wiel 2018, private communication).  The Bridge is well detected in C$^{17}$O gas emission \citep[see also,][]{Favre14}, with the gas following the curvature seen in dust continuum \citep[van der Wiel et al. submitted; see also,][]{Jacobsen18}.  We note that this velocity dispersion is a factor of two higher than the H$^{13}$CO$^+$ (4-3) value used in \citet{Rao09} for their magnetic field strength calculation.  \citet{Rao09} measured the velocity dispersion from SMA observations at $\sim 2\arcsec$ resolution and mostly represents the kinematics of IRAS 16293A rather than the Bridge. 

For the dispersion in polarization position angles, we use the b-vectors with $\PI/\sPI > 5$ and $I/\sigma_I > 5$ (see Figure \ref{bfield}) to ensure robust detections.  Figure \ref{angles} shows a peaked distribution in polarization angles with a tail toward smaller angles. These smaller angles correspond to a slight rotation in position angle seen along the middle of the Bridge near the depolarization zone (see Figure \ref{bfield}).  If we exclude these twisted vectors and fit a Gaussian to the peak distribution, we find $\sigma_{\theta} \approx 5$\degree.  If we take the standard deviation of the entire distribution, including the twisted vectors, we find $\sigma_{\theta} = 17$\degree.  We consider this full range for the magnetic field strength calculation.

We estimate the Bridge density by measuring its mass and volume.  Figure \ref{slice} shows 85 intensity slices through the Bridge, and we use the average of these intensity slices to determine the flux density and width of the Bridge.  The average slice has a median flux density of 3.4 \mJybeam\ and a FWHM of $0.96$\arcsec\ (134 au).  We also adopt length for the Bridge of 3\arcsec\ (420 au) based on where the filament shows uniform b-vectors.  This length excludes the section of the Bridge near IRAS 16293A that is south of the depolarization zone and the section of the Bridge that curves toward IRAS 16293B.  As a result, our adopted length is smaller than the $636$ au used in previous assessments \citep[][; van der Wiel submitted]{Jacobsen18}.  Using the aforementioned median flux density and Bridge dimensions, we estimate a total flux of 0.13 Jy.  We convert this thermal dust emission to mass and column density following a modified blackbody function,
\begin{eqnarray}
M = \frac{S_{\nu}d^2}{\kappa_{\nu}B(\nu,T)},
\end{eqnarray}
where $S_{\nu}$ is the total flux at 233 GHz, $d$ is the distance, $\kappa_{\nu}$ is the dust and gas mass opacity, and $B(\nu,T)$ is the blackbody distribution for a dust temperature $T$.  We assume a temperature of 30 K \citep{Jacobsen18} and a dust opacity at 233 GHz of  0.01112 \cmg\ at 1.3 mm following the theoretical measurements from \citet[][]{Ossenkopf94} for gas densities of 10$^8$ \vol\ and a gas to dust ratio of 100.  Thus, we find a total mass of 0.026 \Msun\ for the Bridge, a typical column density of $1.1 \times 10^{24}$ \cden, and a typical density of $5.6 \times 10^{8}$ \vol, assuming a mean molecular weight per hydrogen molecule of 2.8 for a cloud of 71\%\ molecular hydrogen gas, 27 \%\ helium, and 2\%\ metals \citep{Kauffmann08}, and that the Bridge can be treated like a cylinder.

Assuming $n = 5.6 \times 10^8$ \vol, $\sigma_v = 0.75$ \kms\ and $\sigma_{\theta} = 5-17\degree$, we find a magnetic field strength of $B = 23-78$ mG.  \citet{Alves12} measured a line-of-sight field strength of 110 mG from Zeeman splitting in a water maser toward IRAS 16293A and \citet{Rao09} measured 4.5 mG from the Davis-Chandrasekhar-Fermi method with coarser resolution (2\arcsec, 280 au) dust polarization observations, assuming the latent magnetic field follows an hourglass morphology.  For \citet{Alves12}, water masers are associated with high-density post-shock gas.  Since flux-frozen magnetic fields scale with density, we should expect higher field strengths from water masers than with a quiescent Bridge of gas between binary stars.   In the case of \citet{Rao09}, they measured a density that was lower than our estimate for the Bridge by roughly an order of magnitude and a velocity dispersion that was a factor of two smaller.  As mentioned above, the Bridge density is highly uncertain and could be overestimated by our assumed temperature and dust opacity.  \citet{Rao09} also measured the magnetic field strength across IRAS 16293A assuming it follows an hourglass shape, whereas we measure the field strength across the Bridge between IRAS 16293A and IRAS 16293B.  There is no reason these measurements should be identical.  We see a lot more structure in the field pattern around IRAS 16293A than what was seen by \citet{Rao09}, and do not attempt to compare it with an hourglass morphology.  

Uncertainties in dust opacity and dust temperature, however, can affect these mass and density measurements by an order of magnitude.    In attempt to better constrain models, van der Wiel et al. (submitted) used radiative transfer models of both dust continuum and dense molecular line tracers for the Bridge.  They treated the Bridge like a curved cylinder of length 636 au that is being heated by two luminous stars on either end \citep[e.g., following][]{Jacobsen18}.  While they find a preferred peak density of 7.5 $\times$ 10$^{8}$ \vol, which is in agreement with our value, they  suggest that the Bridge likely has densities between 4 $\times$ 10$^{4}$ \vol and 3 $\times$ 10$^{7}$ \vol\ based on nondetections of several molecules with critical densities of $\sim 10^8$ \vol.  If we instead adopt their upper limit density estimate of 3 $\times$ 10$^{7}$ \vol, our magnetic field strengths will decrease by a factor of 4.  Nevertheless, the critical densities of molecular species do not directly translate to physical densities \citep{Shirley15}.   \citet{Kauffmann17} showed that HCN (1-0) can be well detected in clouds at physical densities much lower than the transition critical value.   The presence of the molecular species in the gas phase and the excitation conditions are both significant to detections, and the chemistry of the Bridge is not yet well constrained.  Thus, we use only our estimated density of $5.6 \times 10^{8}$ \vol\ for the magnetic field measurement at this time.


\section{Discussion}\label{discussion}

\subsection{The Bridge and Streamers}

Figure \ref{bfield} shows that the inferred plane-of-sky magnetic field morphology is aligned mainly parallel with the Bridge and the two streamers seen in IRAS 16293 (see Section \ref{parallel}).  We can use this preferred orientation to investigate both the origin of these dense ($\sim 10^8$ \vol)  filamentary structures as well as their impact on the star formation process in IRAS 16293A and IRAS 16293B.  For simplicity, we focus on the Bridge material, assuming the A-streamer and B-streamer have similar properties to it.

\subsubsection{Parallel Magnetic Fields in Larger-Scale Filaments}\label{previous}

 In this section, we compare the Bridge to observations of larger-scale filaments and filamentary clouds that have parallel field orientations.   \citet{PlanckB15} used thermal dust emission from \emph{Planck} to characterize the field structure for the largest sample of elongated clouds to date.  They found that that the clouds with column densities $> 5 \times 10^{21}$ \cden\ had fields perpendicular to the cloud elongation and that the clouds with column densities $< 5 \times 10^{21}$ \cden\ had parallel fields.  Similar field orientations are also seen from near-infrared polarization measurements, which have superior angular resolution to the \emph{Planck} analyses.  In the near-infrared studies, low-density ($\sim 10^{21}$ \cden) striations have magnetic fields parallel to their elongation, and denser ($\sim 10^{22}$ \cden) filaments have fields that are perpendicular \citep[e.g.,][]{Goldsmith08,Chapman11, Palmeirim13, FrancoAlves15, Panopoulou16, Santos16}.

More recently, \citet{Monsch18} identified a parallel magnetic field toward a dense filament, F2, in the OMC1 region of the Orion molecular cloud.  They used \ammonia\ observations from the VLA at 0.01 pc (2000 au) scales to characterize the physical properties of F2 and dust polarization observations on 0.03 pc (5000 au) scales from \citet{Pattle17} to determine its plane-of-sky magnetic field orientation.  Unlike the aforementioned striations that have parallel field orientations, F2 is very dense.  \citet{Monsch18} estimate its column density at $10^{22.5-23}$ \cden, which is at least an order of magnitude higher than the column density threshold identified by \emph{Planck}.  The inferred magnetic field direction is well aligned with F2.  The plane-of-sky magnetic field is typically within 11\degree\ of the filament orientation across a length of 0.45 pc (4\arcmin).   

The F2 filament is also kinematically active.  It has a velocity gradient across its long axis, and the filament also appears to be aligned with the energetic outflow emanating from Orion BN/KL \citep{Monsch18}.  Recently, \citet{Gomez18} showed that accretion along filaments can bend magnetic field lines  when the flow along the filament is very high \citep[see also,][]{Seifried13, ChenOstriker14}.  The resulting field structure would be ``U-shaped'', because the field is dragged along with the gas.  Since the polarization observations of F2 do not spatially resolve the filament, the U-shape may not be itself resolved.  

F2 has similar properties to the Bridge.  Both share a parallel field morphology and the Bridge is only an order of magnitude smaller in scale and an order of magnitude higher in density (see Section \ref{strength}) than F2.  The Bridge also does not show a U-shape field, even though its filamentary structure is spatially resolved.  Unlike F2, however, the Bridge appears to be kinematically quiescent.  It is distinct from the protostellar outflows and its gas motions are consistent with no line-of-sight velocity gradient (van der Wiel et al. submitted).  Molecular spectra, however, do not trace transverse motions.  If the Bridge is primarily in the plane-of-the-sky, then we could be missing its bulk gas flow.   Indeed, both IRAS 16293A and IRAS 16293B show spectral features indicative of infall \citep[e.g.,][]{Pineda12}.   Since both stars are accreting material, then the dust could be tracing gas flows infalling onto the protostars rather than outflowing as in F2.  

\subsubsection{Magnetic Fields Around Protostars on Small Scales}\label{previous_small}

The IRAS 16293 dust polarization observations also show a number of similarities to other recent studies that examined dust polarization and magnetic fields in the vicinity of protostars ($\lesssim 1000$ au scales).  For simplicity, we focus on studies that trace magnetic fields in inner envelopes around protostars rather than in disks, where dust polarization is primarily attributed to dust self-scattering \citep[e.g.,][]{Kataoka16, Stephens17, Hull18}.  

Several recent ALMA studies of protostellar envelopes have traced the dust polarization in low-mass and high-mass systems at $\lesssim 1000$ au scales \citep[e.g.,][]{Hull17, Hull17smm1, Cox18, Maury18, Koch18, Kwon18}.  These observations show a wide range of inferred magnetic field morphologies.  B335 \citep{Maury18} and L1448  IRS 2 \citep{Kwon18} have relatively symmetric magnetic fields that are consistent with expectations of a pinched, hourglass magnetic field.  The observations of B335, however, may be complicated by strong dust polarization associated with the outflow cavity walls \citep[see also Serpens SMM1,][]{Hull17smm1}.  \citet{Cox18} observed several protostars in Perseus in dust polarization at 80 au resolution and found mainly ordered polarization structures in their envelopes, although they do not model the inferred magnetic field morphology with hourglass shapes.  \citet{Koch18} observed the W51 high-mass star-forming complex at 1000 au resolution and similarly found distinct magnetic field structures, ranging from streamlined channels to converging zones and cometary shapes.  By contrast, \citet{Hull17} found a complicated and highly distorted magnetic field morphology for Ser-emb-8 at 140 au resolution.   Using magnetohydrodynamic (MHD) simulations, \citet{Hull17} concluded that the magnetic field in Ser-emb-8 must be weak relative to gravity and turbulence, such that the magnetic field itself is being altered by the gas dynamics. 

In the case of parallel magnetic fields within filamentary structures,  \citet{Koch18} argued that gravity will be unopposed and gas will be able to flow freely.   Such discrete gas flows have been seen in recent MHD simulations of turbulent disk formation by \citet{Seifried13} and \citet{Seifried15}.  In these simulations, stars accrete mass and angular momentum anisotropically through one or a few narrow channels on scales of $\sim 100$ au to $\sim 1000$ au.  These accretion channels are expected to be important and account for 50\%\ of the mass accreted onto the star and disk.  Within these channels, the magnetic field will be dragged with the accretion flow, whereas on larger scales outside of the channels, the field structure is expected to be disordered due to (even subsonic) turbulence \citep{Seifried15}.  In Section \ref{toy} we construct a toy model to determine the properties of such a flow. 

\subsubsection{Depolarization} \label{depol}

In this section, we discuss how the depolarization (lower polarization fractions) seen in the Bridge may correspond to changes in dust grain properties or the magnetic field structure.  We see zones of depolarization between the Bridge and B-Streamer and across the Bridge near the midpoint between the two stars (see Figure \ref{bfield}). The zones have a width of at least one beam, are spatially resolved in several areas, and have clear dust emission.  Their Stokes I intensity is $\sim 3$ \mJybeam, which is $> 10$ $\sigma_I$.  Based on this intensity, we estimate a 4$\sPI$ upper limit of 3\%.  This upper limit is comparable to the polarization fractions seen along the edge of the depolarization zones.  Thus, we may simply lack the sensitivity to detect the dust polarization in this region.

Nevertheless, we can conclude that the polarization fractions in these zones are lower than elsewhere in the Bridge.  Lower polarization fractions can arise from less effective grain alignment, disordered magnetic fields, or unresolved magnetic field structure along the line of sight \citep[e.g.,][]{Padoan01,ChoLazarian05,Alves14}. Grain alignment depends on the time scales necessary to align the spinning dust with the magnetic field where turbulence and gas drag can remove this alignment.   Large ($\gtrsim 10\ \um$) dust grains or dust grains with less paramagnetic material are more difficult to align \citep[e.g.,][]{Tazaki17}.  \citet{Jorgensen16} showed a three-color image of IRAS 16293 using dust continuum observations at 0.87 mm, 1.3 mm, and 3 mm from ALMA (see their Figure 3).  This image shows distinct regions of bluer colors in the Bridge that correspond solely with the depolarization zones in Figure \ref{bfield}.  Bluer emission has a steeper spectral index that may indicate higher temperatures or larger values of the dust emissivity index, $\beta$.   Larger values of $\beta$, however, are typically associated with smaller dust grains \citep[e.g.,][]{Ossenkopf94, Ormel11, Testi14}, which should be more easily aligned with the magnetic field.  Therefore, we see no evidence that the depolarization zones are due to changes in the dust grains themselves.   

The depolarization may instead be due to changes in the magnetic field structure itself.   As mentioned above, the bluer emission seen by \citet{Jorgensen16} can also indicate slightly higher dust temperatures.  If the higher dust temperatures coincide with excess turbulence or shocks, then the magnetic field in the depolarization zones could be more disordered.  Disordered fields are less efficient at aligning dust grains and will have lower polarization fractions \citep[e.g.,][]{Hull17, Seifried18}.

Alternatively, the depolarization zones may represent unresolved magnetic field structure caused by the two protostars.   The two stars appear to dominate the gravitational potential.  IRAS 16293A has an estimated mass of $\sim 0.5-1.0$ \Msun\ with IRAS 16293B at $\sim 0.1$ \Msun\ \citep{Caux11, Pineda12, Oya16}.  These masses are roughly an order of magnitude higher than our estimate of the Bridge mass (0.026 \Msun; see Section \ref{strength}).  A flux-frozen magnetic field will therefore be split by the gravitational influence of the two stars.  Indeed, \citet{Rao09} identified an hourglass-shaped field toward IRAS 16293A, and we see hints of a separate pinched field morphology toward the outer edge of IRAS 16293B.  In this case, the depolarization zones may represent the regions where the field is being pulled in opposing directions.  

\subsubsection{Toy Model}\label{toy}

In this section, we construct a simple toy model of the Bridge assuming that IRAS 16293A and IRAS 16293B dominate the mass and gravitational potential of the system (see Section \ref{depol}).  We model the Bridge by a straight cylinder of length $2L = 740$ au and negligible mass.  The adopted length for the cylinder represents the full distance between IRAS 16293A and IRAS 16293B scaled to our assumed distance of 140 pc \citep[][van der Wiel et al. submitted]{Jacobsen18}.   Two stars are placed at either end of the cylinder.  For simplicity, we assume that the stars have the same mass, $M$ and the Bridge is in the plane of the sky.

A point mass along this toy cylinder will feel a gravitational pull from both stars.  If the point mass is displaced a distance $x$ from the center of the cylinder, then the gravitational acceleration it will feel is,
\begin{equation}
g = GM\left[\frac{1}{(L-x)^2} - \frac{1}{(L+x)^2}\right].
\end{equation}
Since the stars dominate the gravitational potential, the point mass should fall toward the nearest star.  As the point mass moves from an initial position $x_0$ to a new position $x$ in the cylinder, it will gain kinetic energy corresponding to the change in potential energy.  The velocity flow from this gain in kinetic energy is,
\begin{equation}
v_{flow}^2 = v_0^2 + \frac{4GM}{L}\left[\frac{1}{1-(x/L)^2} - \frac{1}{1-(x_0/L)^2}\right],\label{vflow_eq}
\end{equation}
where $v_0$ is the initial velocity of the point mass at position $x_0$.  The flow from rest can be written as $v_{flow} = a\sqrt{GM/L}$, where $a$ is a constant that depends on the change in position.   In general, $a$ is of order unity.  A point mass that moves from rest near the center of the cylinder to a position halfway to the star will have $a = 1.2$, with values that range from $a=0.5$ for $x = L/4$ to $a = 2.3$ for $x=3L/4$.   Thus, we can expect flow rates of $v_{flow} = 0.5-1.5$ \kms\ for star masses between 0.1 and 1.0 \Msun.  

We can compare the flow rate in our toy model with the expected Alfv\'{e}n speed.  The Alfv\'{e}n speed is comparable to the sound speed in a magnetized medium and is given by,
\begin{equation}
v_A= \frac{B}{\sqrt{4\pi\rho}},
\end{equation}
where $B$ is the magnetic field strength and $\rho = \mu m_Hn$ is the gas density.  Using our values from Section \ref{strength}, we find $v_A = 1.2-4$ \kms, which is of similar order to the flow speed, although there is a large range in both cases.   Trans-Alfv\'{e}nic or moderately sub-Alfv\'{e}nic magnetic field may be strong enough to regulate the flow itself \citep[e.g.,][]{Burkhart09, Hull17}.  This regulation could explain the relatively narrow range of velocities seen across the Bridge by van der Wiel et al. (submitted).  We note that the expected gas flow is comparable to the the magnitude of the velocity range seen in the Bridge.  If the Bridge is mainly in the plane of the sky, this flow would not be easily detected (see Section \ref{previous}). 

We can also compute the time scale for the gas to flow to each star.  The gas flow time is given by $\tau_{flow} = L/v_{flow} = b\sqrt{L^3/GM}$, where $b$ is a numerical constant of order unity from integrating the velocity function in Equation \ref{vflow_eq}.  For our toy cylinder, $\tau_{flow} \sim 10^3$ yr.  This time scale is much shorter than the expected ages for these stars.  Both IRAS 16293A and IRAS 16293B are considered Class 0 objects, although they could have different ages from each other.  The typical Class 0 lifetime is $\sim 0.2$ Myr \citep{Dunham15}, which is two orders of magnitude larger than the estimated flow rate.  Unless the gas in the Bridge is replenished, we would expect it to be entirely depleted.  

van der Wiel et al. proposed that the Bridge is the remnant of the filamentary core that initially fragmented to form IRAS 16293A and IRAS 16293B.  If that is the case, then the Bridge must be accreting new material. In their MHD simulations, \citet{Gomez18} found that even after star formation occurs, filaments could accrete gas directly from their surroundings.  The infalling gas only curved away from  the filament when in close proximity to a dense star-forming hub.  The magnetic field structure in these cases should be perpendicular to the filament to allow the gas to flow directly to it and then bend to be parallel to follow the direction of the flow.  Figure \ref{bfield}, however, shows uniform magnetic fields that are parallel, not perpendicular, to the Bridge.  There is some evidence of the field curving on the top edge of the Bridge, toward the depolarization zone, but the orientation is still mainly parallel.   Since gas cannot flow across field lines easily, we cannot conclude that the Bridge is accreting new material.  Either this accretion is not taking place or gas is infalling onto the Bridge on larger scales than what our ALMA data can recover.   Additional, sensitive observations that trace the magnetic field structure over larger spatial scales may be able to rule out or confirm accretion onto the Bridge.  

Alternatively, the Bridge may not be a natal structure and may instead be transient.  Numerous MHD simulations of collapsing cores have produced stars with bars and spiral structures through tidal forces, gravitational instabilities, or fragmentation \citep[e.g.,][]{PriceBate07, Machida08, HennebelleTeyssier08, Burzle11, Commercon11, Joos13, BossKeiser13, Masson16, Offner16}, and the accretion channels discussed in Section \ref{previous_small} can vary on a few kiloyear timescales \citep{Seifried15}.  Many of the aforementioned simulations also produce material bridging young binary stars, much like what we see in IRAS 16293, and these structures can change in size and shape very quickly in a freefall timescale.  If IRAS 16293A and IRAS 16293 are gravitationally bound to each other, their orbital period from Kepler's laws is $P = 6.3\sqrt{L^3/G(M_1 + M_2)}$.  This orbital period is of similar order to the flow time scale assuming the stars have the same mass.   Therefore, the Bridge may instead be a tidal feature of IRAS 16293 that will eventually dissipate or transform into a different structure in the next few thousand years.

\subsection{Polarization of IRAS 16293B}\label{polB}

\begin{figure}[h!]
\includegraphics[width=0.475\textwidth,trim=1pt 1pt 1pt 1pt,clip=true]{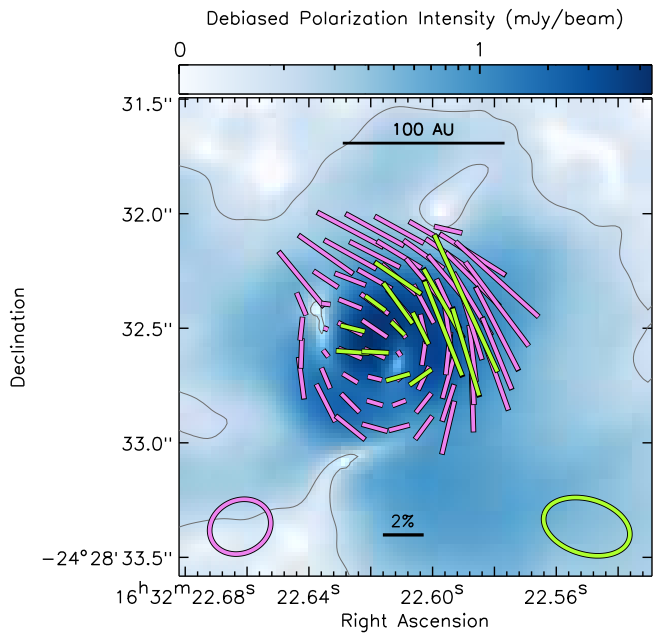}
\caption{Comparison of ALMA 1.3 mm polarization (purple e-vectors) with VLA 6.9 mm polarization (green e-vectors) from \citet{Liu18} toward IRAS 16293B.  We use the Q-band e-vectors with $\PF < 10$\%\ from the $\mbox{\texttt{robust}} = 0$ results for the Q-band dust polarization for this comparison.  For clarity, only those vectors within $\sim 0.5$\arcsec\ of the map center are shown.  The background image shows the debiased polarized intensities from Figure \ref{polarization}.  The ALMA and VLA beams are shown in the lower-left and lower-right corners, respectively.  \label{qband}}
\end{figure}

In this section, we examine the dust polarization of IRAS 16293B specifically.  \cite{Rao14} observed IRAS 16293 in 878 \um\ dust polarization with the SMA at $\sim 0.6$\arcsec\ (84 au) resolution, and more recently, \citet{Liu18} presented 6.9 mm polarization observations of IRAS 16293B from the Karl G. Jansky Very Large Array (VLA) at similar resolution to the present ALMA 1.3 mm observations.  Figure \ref{qband} compares our 1.3 mm ALMA dust polarization e-vectors (violet) with the 6.9 mm VLA e-vectors (green) form \citet{Liu18}.   The two datasets show remarkable agreement with each other in terms of polarization fraction and position angle.  The inferred magnetic field structure in Figure \ref{bfield} also shows excellent agreement with the 878 \um\ polarization measurements from \citet{Rao14}. 

\citet{Rao14} modeled their 878 \um\ dust polarization with a toroidal-like magnetic field.  Nevertheless, recent ALMA dust polarization studies of protostellar disks have shown that polarization signatures can arise from alternative mechanisms other than magnetic grain alignment.  These mechanisms include self-scattering of dust grains in an anisotropic radiation field \citep[e.g.,][]{Kataoka15, Yang16, Pohl16} or grain alignment from the radiative torques themselves \citep[e.g.,][]{LazarianHoang07, Tazaki17}.  The polarization signature from self-scattering and radiative grain alignment can be predicted using the geometric properties of the disk and its gradient of radiation.  Features such as inclination and gaps will affect the observed gradient of radiation and the predicted polarization pattern.  Multiwavelength observations can help disentangle these effects because we expect self-scattering to dominate the polarization signature at shorter wavelengths, when disks are optically thick, and radiative grain alignment at longer wavelengths, which primarily trace larger ($\gtrsim 50$ \um) dust grains \citep[see also,][]{Kataoka17, Stephens17, Yang17}.  
 
For IRAS 16293B, we see consistent polarization patterns from 878 \um\ \citep{Rao09}, 1.3 mm (this work), and 6.9 mm \citep{Liu18}, which suggests that there is a single polarization mechanism across these bands.   \citet{Alves18} similarly found consistent polarization observations at 870 \um, 1.3 mm, and 3 mm for the circumbinary disk of BHB07-11.  They concluded that the polarization pattern of BHB07-11 was inconsistent with either self-scattering or radiative grain alignment, and they instead suggested grain alignment from mainly poloidal magnetic fields.  Qualitatively, the polarization signature toward IRAS 16293B is reminiscent of dust self-scattering in a face-on disk \citep[e.g.,][]{Kataoka15, Yang16}.  Figure \ref{qband} shows an ordered structure that is azimuthal at its periphery and uniform within one beam of its center.  Although an azimuthal polarization morphology is expected in the case of radiative grain alignment \citep{Tazaki17}, this mechanism does not predict uniform polarization toward the center, although the latter is not well resolved.   Moreover, the dust emission, especially at 878 \um\ and 1.3 mm, should be optically thick and therefore more likely to trace dust self-scattering than any other mechanism.

The similarity between the 1.3 mm polarization and 6.9 mm polarization, however, is more surprising because it suggests that even the radio frequency data are also tracing dust scattering, even though such emission is usually considered optically thin. Nevertheless,  \citet{Yang16_iras4a} argued that dust polarization observations of IRAS 4A at 8 mm from \citet{Cox15} can be entirely attributed to dust self-scattering, so there are other cases with self-scattering at radio frequencies.  IRAS 16293B in particular has a spectral slope of $S_{\nu} \sim \nu^2$ even down to 8 GHz frequencies \citep{Chandler05}.  Such spectral slopes at radio frequencies can be attributed to either optically thick emission (e.g., the dust emits as a perfect blackbody rather than a modified blackbody) or to very large dust grains that are $> 1$ mm \citep[e.g.,][]{Natta07, Kwon09, Pinte10, Ricci10, Testi14}.  Both of these explanations can produce a self-scattering signature at 6.9 mm.  As stated above, dust scattering is most dominant in optically thick disks.  In addition, polarized self-scattering is highly dependent on the grain size distribution, and a detection at 6.9 mm implies grain sizes of $\sim 200-2000$ \um\ \citep{Kataoka15}, which agrees well with the observed spectral slope.  We note, however, that this level of grain growth is higher than what has been previously identified in this region \citep[e.g.,][]{Chandler05, Harada17}.  Further analysis with dust scattering models using the multiwavelength observations is still necessary to determine the maximum grain sizes.   If verified by models, these observations would suggest that very large dust grains can form quickly in young disks.

\section{Summary and Conclusions}\label{summary}

We present new ALMA 1.3 mm polarization observations for IRAS 16293.  These data spatially resolve the polarization structure around the IRAS 16293A and IRAS 16293B protostars that was previously seen in lower resolution data.    Our main results are:

\begin{enumerate}

\item We find extensive, highly ordered dust polarization around IRAS 16293A and IRAS 16293B.  We also find ordered dust polarization across the Bridge of dust between the two protostars. 

\item The inferred magnetic field structure of IRAS 16293 is more chaotic than what was seen in previous lower resolution observations by \citet{Rao09}.  Disordered magnetic fields have been seen in recent observational and theoretic work and may indicate that turbulence is affecting magnetic fields on small scales. There are hints of separate pinched field morphologies for both IRAS 16293A and IRAS 16293B, but we lack short $uv$ spacings to fully recover the structure of the latter.

\item We also find that the inferred magnetic field morphology in the Bridge, A-Streamer, and B-streamer is  aligned  with the long axes of these filamentary structures.   Parallel magnetic fields have not been broadly seen toward filamentary structures on these scales ($\lesssim 400$ au) before and may represent accretion channels discussed in simulations.

\item We characterize the Bridge properties and find a mass of 0.026 \Msun, a density of $5.6 \times 10^{8}$, in broad agreement with previous measurements from radiative transfer models \citep[van der Wiel et al. submitted][]{Jacobsen18}.  We also measure a field strength of $23-78$ mG using the Davis-Chandrasekhar-Fermi method.  

\item This inferred magnetic field for IRAS 16293A and IRAS 16293B is separated by zones of depolarization near the midpoint between both stars.  We find that this depolarization zone is correlated with bluer dust emission as seen by \citet{Jorgensen16} and may indicate that the region has higher dust temperatures.   We suggest that the depolarization represents unresolved magnetic field structure due to a transition in the field being dominated by IRAS 16293A to being dominated by IRAS 16293B, or to a more disordered field due to higher degrees of turbulence.  

\item We construct a toy model for the Bridge to assess its formation and evolution.  We find that expected gas flow rate of $\sim 1$ \kms\ to each star is comparable to the Alfv\'{e}n speed of the material, which suggests that the magnetic fields could regulate the gas flow in the filamentary structures.  We also find that the time scale for gas to flow out of the Bridge is $\sim 10^3$ yr.  This time scale is much shorter than the Class 0 lifetime.  If the Bridge is part of the original natal filamentary system that formed the protostars, it must accrete new material directly from its surroundings.  We see no evidence of this accretion in our observations, but we may lack the small spatial frequencies to recover it.  Alternatively, the Bridge may be a transient structure produced by the gravitational interplay between the forming stars.

\item We propose that the dust polarization seen toward the IRAS 16293B disk is due to dust self-scattering and not magnetic grain alignment.  The morphology of the disk polarization is consistent with predictions of dust self-scattering for a face-on disk and dust self-scattering is the favored mechanism for optically thick disks \citep{Yang17}.   \citet{Liu18} find similar polarization fractions and position angles in this disk at 6.9 mm.  If both the 1.3 mm and 6.9 mm polarization observations are due to dust self-scattering, then IRAS 16293B has formed dust grains sizes with sizes between $\sim 200 - 2000$ \um.
 
\end{enumerate}

These observations highlight the benefit of high angular resolution dust polarization studies.  Previous lower resolution data did not capture the relationship between the inferred plane-of-sky magnetic field direction and the elongated, filamentary dust features in IRAS 16293.  The parallel alignment between the inferred magnetic field and the filamentary structures around the young stars is a unique feature.  Future theoretical work will need to understand the origin of this parallel magnetic field structure and the impact it may have on the evolution of the young stars.  


\vspace{1cm}
\begin{acknowledgements}
We thank the anonymous referee for comments that helped improve the discussion.  The authors also acknowledge the NAASC and EU-ARC for support with the ALMA observations and data processing.  The authors thank Shantanu Basu, Sebastien Fromang, Lee Hartmann, Patrick Hennebelle, Zhi-Yun Li, Stella Offner, and Anna Rosen for insightful theoretical discussions on magnetic fields in protostellar systems and filaments, and Ted Bergin, Jes J{\o}rgensen, and Matthijs van der Wiel for valuable discussions on the kinematic and physical properties of the dust and gas in the Bridge.  S.I.S. acknowledges the support for this work provided by NASA through Hubble Fellowship grant No. HST-HF2-51381.001-A awarded by the Space Telescope Science Institute, which is operated by the Association of Universities for Research in Astronomy, Inc., for NASA, under contract No. NAS 5-26555.  W.K. was supported by Basic Science Research Program through the National Foundation of Korea (grant No. NRF-2016R1C1B2013642). L.W.L. acknowledges the support of NASA/JPL through the grant No. 1594465.
This paper makes use of the following ALMA data: ADS/JAO.ALMA\#2015.1.01112.S. ALMA is a partnership of ESO (representing its member states), NSF (USA) and NINS (Japan), together with NRC (Canada), MOST and ASIAA (Taiwan), and KASI (Republic of Korea), in cooperation with the Republic of Chile. The Joint ALMA Observatory is operated by ESO, AUI/NRAO and NAOJ.  The National Radio Astronomy Observatory is a facility of the National Science Foundation operated under cooperative agreement by Associated Universities, Inc.

\end{acknowledgements}

\bibliographystyle{apj}

\begin{thebibliography}{}
\expandafter\ifx\csname natexlab\endcsname\relax\def\natexlab#1{#1}\fi

\bibitem[{{Alves} {et~al.}(2014){Alves}, {Frau}, {Girart}, {Franco}, {Santos},
  \& {Wiesemeyer}}]{Alves14}
{Alves}, F.~O., {Frau}, P., {Girart}, J.~M., {et~al.} 2014, \aap, 569, L1

\bibitem[{{Alves} {et~al.}(2012){Alves}, {Vlemmings}, {Girart}, \&
  {Torrelles}}]{Alves12}
{Alves}, F.~O., {Vlemmings}, W.~H.~T., {Girart}, J.~M., \& {Torrelles}, J.~M.
  2012, \aap, 542, A14

\bibitem[{{Alves} {et~al.}(2018){Alves}, {Girart}, {Padovani}, {Galli},
  {Franco}, {Caselli}, {Vlemmings}, {Zhang}, \& {Wiesemeyer}}]{Alves18}
{Alves}, F.~O., {Girart}, J.~M., {Padovani}, M., {et~al.} 2018, \aap, 616, A56

\bibitem[{{Anderson} \& {Darling}(1952)}]{AndersonDarling52}
{Anderson}, T.~W., \& {Darling}, D.~A. 1952, Annals of Mathematical Statistics,
  23, 193

\bibitem[{{Anderson} \& {Darling}(1954)}]{AndersonDarling54}
---. 1954, Journal of the American Statistical Association, 49, 765

\bibitem[{{Andersson} {et~al.}(2015){Andersson}, {Lazarian}, \&
  {Vaillancourt}}]{Andersson15}
{Andersson}, B.-G., {Lazarian}, A., \& {Vaillancourt}, J.~E. 2015, \araa, 53,
  501

\bibitem[{{Bacciotti} {et~al.}(2018){Bacciotti}, {Girart}, {Padovani}, {Podio},
  {Paladino}, {Testi}, {Bianchi}, {Galli}, {Codella}, {Coffey}, {Favre}, \&
  {Fedele}}]{Bacciotti18}
{Bacciotti}, F., {Girart}, J.~M., {Padovani}, M., {et~al.} 2018, \apjl, 865,
  L12

\bibitem[{{Blandford} \& {Payne}(1982)}]{BlandfordPayne82}
{Blandford}, R.~D., \& {Payne}, D.~G. 1982, \mnras, 199, 883

\bibitem[{{Boss} \& {Keiser}(2013)}]{BossKeiser13}
{Boss}, A.~P., \& {Keiser}, S.~A. 2013, \apj, 764, 136

\bibitem[{{Burkhart} {et~al.}(2009){Burkhart}, {Falceta-Gon{\c c}alves},
  {Kowal}, \& {Lazarian}}]{Burkhart09}
{Burkhart}, B., {Falceta-Gon{\c c}alves}, D., {Kowal}, G., \& {Lazarian}, A.
  2009, \apj, 693, 250

\bibitem[{{B{\"u}rzle} {et~al.}(2011){B{\"u}rzle}, {Clark}, {Stasyszyn},
  {Greif}, {Dolag}, {Klessen}, \& {Nielaba}}]{Burzle11}
{B{\"u}rzle}, F., {Clark}, P.~C., {Stasyszyn}, F., {et~al.} 2011, \mnras, 412,
  171

\bibitem[{{Caux} {et~al.}(2011){Caux}, {Kahane}, {Castets}, {Coutens},
  {Ceccarelli}, {Bacmann}, {Bisschop}, {Bottinelli}, {Comito}, {Helmich},
  {Lefloch}, {Parise}, {Schilke}, {Tielens}, {van Dishoeck}, {Vastel},
  {Wakelam}, \& {Walters}}]{Caux11}
{Caux}, E., {Kahane}, C., {Castets}, A., {et~al.} 2011, \aap, 532, A23

\bibitem[{{Chandler} {et~al.}(2005){Chandler}, {Brogan}, {Shirley}, \&
  {Loinard}}]{Chandler05}
{Chandler}, C.~J., {Brogan}, C.~L., {Shirley}, Y.~L., \& {Loinard}, L. 2005,
  \apj, 632, 371

\bibitem[{{Chandrasekhar} \& {Fermi}(1953)}]{ChandrasekharFermi53}
{Chandrasekhar}, S., \& {Fermi}, E. 1953, \apj, 118, 113

\bibitem[{{Chapman} {et~al.}(2011){Chapman}, {Goldsmith}, {Pineda}, {Clemens},
  {Li}, \& {Kr{\v c}o}}]{Chapman11}
{Chapman}, N.~L., {Goldsmith}, P.~F., {Pineda}, J.~L., {et~al.} 2011, \apj,
  741, 21

\bibitem[{{Chen} \& {Ostriker}(2014)}]{ChenOstriker14}
{Chen}, C.-Y., \& {Ostriker}, E.~C. 2014, \apj, 785, 69

\bibitem[{{Chen} {et~al.}(2013){Chen}, {Arce}, {Zhang}, {Bourke}, {Launhardt},
  {J{\o}rgensen}, {Lee}, {Foster}, {Dunham}, {Pineda}, \& {Henning}}]{Chen13}
{Chen}, X., {Arce}, H.~G., {Zhang}, Q., {et~al.} 2013, \apj, 768, 110

\bibitem[{{Cho} \& {Lazarian}(2005)}]{ChoLazarian05}
{Cho}, J., \& {Lazarian}, A. 2005, \apj, 631, 361

\bibitem[{{Cho} \& {Lazarian}(2007)}]{ChoLazarian07}
---. 2007, \apj, 669, 1085

\bibitem[{{Commer{\c c}on} {et~al.}(2011){Commer{\c c}on}, {Hennebelle}, \&
  {Henning}}]{Commercon11}
{Commer{\c c}on}, B., {Hennebelle}, P., \& {Henning}, T. 2011, \apjl, 742, L9

\bibitem[{{Cox} {et~al.}(2018){Cox}, {Harris}, {Looney}, {Li}, {Yang}, {Tobin},
  \& {Stephens}}]{Cox18}
{Cox}, E.~G., {Harris}, R.~J., {Looney}, L.~W., {et~al.} 2018, \apj, 855, 92

\bibitem[{{Cox} {et~al.}(2015){Cox}, {Harris}, {Looney}, {Segura-Cox}, {Tobin},
  {Li}, {Tychoniec}, {Chandler}, {Dunham}, {Kratter}, {Melis}, {Perez}, \&
  {Sadavoy}}]{Cox15}
---. 2015, \apjl, 814, L28

\bibitem[{{Crutcher}(2012)}]{Crutcher12}
{Crutcher}, R.~M. 2012, \araa, 50, 29

\bibitem[{{Crutcher} {et~al.}(2004){Crutcher}, {Nutter}, {Ward-Thompson}, \&
  {Kirk}}]{Crutcher04}
{Crutcher}, R.~M., {Nutter}, D.~J., {Ward-Thompson}, D., \& {Kirk}, J.~M. 2004,
  \apj, 600, 279

\bibitem[{Davis(1951)}]{Davis51}
Davis, L. 1951, Phys. Rev., 81, 890

\bibitem[{{Dolginov} \& {Mitrofanov}(1976)}]{DolginovMitrofanov76}
{Dolginov}, A.~Z., \& {Mitrofanov}, I.~G. 1976, \apss, 43, 291

\bibitem[{{Dunham} {et~al.}(2015){Dunham}, {Allen}, {Evans},
  {Broekhoven-Fiene}, {Cieza}, {Di Francesco}, {Gutermuth}, {Harvey},
  {Hatchell}, {Heiderman}, {Huard}, {Johnstone}, {Kirk}, {Matthews}, {Miller},
  {Peterson}, \& {Young}}]{Dunham15}
{Dunham}, M.~M., {Allen}, L.~E., {Evans}, II, N.~J., {et~al.} 2015, \apjs, 220,
  11

\bibitem[{{Dzib} {et~al.}(2018){Dzib}, {Ortiz-Le{\'o}n},
  {Hern{\'a}ndez-G{\'o}mez}, {Loinard}, {Mioduszewski}, {Claussen}, {Menten},
  {Caux}, \& {Sanna}}]{Dzib18}
{Dzib}, S.~A., {Ortiz-Le{\'o}n}, G.~N., {Hern{\'a}ndez-G{\'o}mez}, A., {et~al.}
  2018, \aap, 614, A20

\bibitem[{{Favre} {et~al.}(2014){Favre}, {J{\o}rgensen}, {Field}, {Brinch},
  {Bisschop}, {Bourke}, {Hogerheijde}, \& {Frieswijk}}]{Favre14}
{Favre}, C., {J{\o}rgensen}, J.~K., {Field}, D., {et~al.} 2014, \apj, 790, 55

\bibitem[{{Field}(1965)}]{Field65}
{Field}, G.~B. 1965, \apj, 142, 531

\bibitem[{{Franco} \& {Alves}(2015)}]{FrancoAlves15}
{Franco}, G.~A.~P., \& {Alves}, F.~O. 2015, \apj, 807, 5

\bibitem[{{Galametz} {et~al.}(2018){Galametz}, {Maury}, {Girart}, {Rao},
  {Zhang}, {Gaudel}, {Valdivia}, {Keto}, \& {Lai}}]{Galametz18}
{Galametz}, M., {Maury}, A., {Girart}, J.~M., {et~al.} 2018, \aap, 616, A139

\bibitem[{{Galli} \& {Shu}(1993)}]{GalliShu93}
{Galli}, D., \& {Shu}, F.~H. 1993, \apj, 417, 220

\bibitem[{{Girart} {et~al.}(2006){Girart}, {Rao}, \& {Marrone}}]{Girart06}
{Girart}, J.~M., {Rao}, R., \& {Marrone}, D.~P. 2006, Science, 313, 812

\bibitem[{{Goldsmith} {et~al.}(2008){Goldsmith}, {Heyer}, {Narayanan}, {Snell},
  {Li}, \& {Brunt}}]{Goldsmith08}
{Goldsmith}, P.~F., {Heyer}, M., {Narayanan}, G., {et~al.} 2008, \apj, 680, 428

\bibitem[{{G{\'o}mez} {et~al.}(2018){G{\'o}mez}, {V{\'a}zquez-Semadeni}, \&
  {Zamora-Avil{\'e}s}}]{Gomez18}
{G{\'o}mez}, G.~C., {V{\'a}zquez-Semadeni}, E., \& {Zamora-Avil{\'e}s}, M.
  2018, \mnras, 480, 2939

\bibitem[{{Harada} {et~al.}(2017){Harada}, {Hasegawa}, {Aikawa}, {Hirashita},
  {Liu}, \& {Hirano}}]{Harada17}
{Harada}, N., {Hasegawa}, Y., {Aikawa}, Y., {et~al.} 2017, \apj, 837, 78

\bibitem[{{Harris} {et~al.}(2018){Harris}, {Cox}, {Looney}, {Li}, {Yang},
  {Fern{\'a}ndez-L{\'o}pez}, {Kwon}, {Sadavoy}, {Segura-Cox}, {Stephens}, \&
  {Tobin}}]{Harris18}
{Harris}, R.~J., {Cox}, E.~G., {Looney}, L.~W., {et~al.} 2018, \apj, 861, 91

\bibitem[{{Hennebelle} \& {Chabrier}(2013)}]{Hennebelle13}
{Hennebelle}, P., \& {Chabrier}, G. 2013, \apj, 770, 150

\bibitem[{{Hennebelle} \& {Teyssier}(2008)}]{HennebelleTeyssier08}
{Hennebelle}, P., \& {Teyssier}, R. 2008, \aap, 477, 25

\bibitem[{{Hoang} \& {Lazarian}(2008)}]{HoangLazarian08}
{Hoang}, T., \& {Lazarian}, A. 2008, \mnras, 388, 117

\bibitem[{{Hou} {et~al.}(2009){Hou}, {Parker}, {Harris}, \& {Wilman}}]{Hou09}
{Hou}, A., {Parker}, L.~C., {Harris}, W.~E., \& {Wilman}, D.~J. 2009, \apj,
  702, 1199

\bibitem[{{Hull} {et~al.}(2014){Hull}, {Plambeck}, {Kwon}, {Bower},
  {Carpenter}, {Crutcher}, {Fiege}, {Franzmann}, {Hakobian}, {Heiles}, {Houde},
  {Hughes}, {Lamb}, {Looney}, {Marrone}, {Matthews}, {Pillai}, {Pound},
  {Rahman}, {Sandell}, {Stephens}, {Tobin}, {Vaillancourt}, {Volgenau}, \&
  {Wright}}]{Hull14}
{Hull}, C.~L.~H., {Plambeck}, R.~L., {Kwon}, W., {et~al.} 2014, \apjs, 213, 13

\bibitem[{{Hull} {et~al.}(2017{\natexlab{a}}){Hull}, {Girart}, {Tychoniec},
  {Rao}, {Cort{\'e}s}, {Pokhrel}, {Zhang}, {Houde}, {Dunham}, {Kristensen},
  {Lai}, {Li}, \& {Plambeck}}]{Hull17smm1}
{Hull}, C.~L.~H., {Girart}, J.~M., {Tychoniec}, {\L}., {et~al.}
  2017{\natexlab{a}}, \apj, 847, 92

\bibitem[{{Hull} {et~al.}(2017{\natexlab{b}}){Hull}, {Mocz}, {Burkhart},
  {Goodman}, {Girart}, {Cort{\'e}s}, {Hernquist}, {Springel}, {Li}, \&
  {Lai}}]{Hull17}
{Hull}, C.~L.~H., {Mocz}, P., {Burkhart}, B., {et~al.} 2017{\natexlab{b}},
  \apjl, 842, L9

\bibitem[{{Hull} {et~al.}(2018){Hull}, {Yang}, {Li}, {Kataoka}, {Stephens},
  {Andrews}, {Bai}, {Cleeves}, {Hughes}, {Looney}, {P{\'e}rez}, \&
  {Wilner}}]{Hull18}
{Hull}, C.~L.~H., {Yang}, H., {Li}, Z.-Y., {et~al.} 2018, \apj, 860, 82

\bibitem[{{Imai} {et~al.}(2007){Imai}, {Nakashima}, {Bushimata}, {Choi},
  {Hirota}, {Honma}, {Horiai}, {Inomata}, {Iwadate}, {Jike}, {Kameya},
  {Kamohara}, {Kan-Ya}, {Kawaguchi}, {Kijima}, {Kobayashi}, {Kuji}, {Kurayama},
  {Manabe}, {Miyaji}, {Nagayama}, {Nakagawa}, {Oh}, {Omodaka}, {Oyama},
  {Sakai}, {Sakakibara}, {Sato}, {Sasao}, {Shibata}, {Shimizu}, {Shintani},
  {Sofue}, {Sora}, {Suda}, {Tamura}, {Tsushima}, {Ueno}, \&
  {Yamashita}}]{Imai07}
{Imai}, H., {Nakashima}, K., {Bushimata}, T., {et~al.} 2007, \pasj, 59, 1107

\bibitem[{{Jacobsen} {et~al.}(2018){Jacobsen}, {J{\o}rgensen}, {van der Wiel},
  {Calcutt}, {Bourke}, {Brinch}, {Coutens}, {Drozdovskaya}, {Kristensen},
  {M{\"u}ller}, \& {Wampfler}}]{Jacobsen18}
{Jacobsen}, S.~K., {J{\o}rgensen}, J.~K., {van der Wiel}, M.~H.~D., {et~al.}
  2018, \aap, 612, A72

\bibitem[{{Joos} {et~al.}(2013){Joos}, {Hennebelle}, {Ciardi}, \&
  {Fromang}}]{Joos13}
{Joos}, M., {Hennebelle}, P., {Ciardi}, A., \& {Fromang}, S. 2013, \aap, 554,
  A17

\bibitem[{{J{\o}rgensen} {et~al.}(2011){J{\o}rgensen}, {Bourke}, {Nguyen
  Luong}, \& {Takakuwa}}]{Jorgensen11}
{J{\o}rgensen}, J.~K., {Bourke}, T.~L., {Nguyen Luong}, Q., \& {Takakuwa}, S.
  2011, \aap, 534, A100

\bibitem[{{J{\o}rgensen} {et~al.}(2016){J{\o}rgensen}, {van der Wiel},
  {Coutens}, {Lykke}, {M{\"u}ller}, {van Dishoeck}, {Calcutt}, {Bjerkeli},
  {Bourke}, {Drozdovskaya}, {Favre}, {Fayolle}, {Garrod}, {Jacobsen},
  {{\"O}berg}, {Persson}, \& {Wampfler}}]{Jorgensen16}
{J{\o}rgensen}, J.~K., {van der Wiel}, M.~H.~D., {Coutens}, A., {et~al.} 2016,
  \aap, 595, A117

\bibitem[{{Kandori} {et~al.}(2017){Kandori}, {Tamura}, {Kusakabe}, {Nakajima},
  {Kwon}, {Nagayama}, {Nagata}, {Tomisaka}, \& {Tatematsu}}]{Kandori17}
{Kandori}, R., {Tamura}, M., {Kusakabe}, N., {et~al.} 2017, \apj, 845, 32

\bibitem[{{Kataoka} {et~al.}(2016{\natexlab{a}}){Kataoka}, {Muto}, {Momose},
  {Tsukagoshi}, \& {Dullemond}}]{Kataoka16}
{Kataoka}, A., {Muto}, T., {Momose}, M., {Tsukagoshi}, T., \& {Dullemond},
  C.~P. 2016{\natexlab{a}}, \apj, 820, 54

\bibitem[{{Kataoka} {et~al.}(2017){Kataoka}, {Tsukagoshi}, {Pohl}, {Muto},
  {Nagai}, {Stephens}, {Tomisaka}, \& {Momose}}]{Kataoka17}
{Kataoka}, A., {Tsukagoshi}, T., {Pohl}, A., {et~al.} 2017, \apjl, 844, L5

\bibitem[{{Kataoka} {et~al.}(2015){Kataoka}, {Muto}, {Momose}, {Tsukagoshi},
  {Fukagawa}, {Shibai}, {Hanawa}, {Murakawa}, \& {Dullemond}}]{Kataoka15}
{Kataoka}, A., {Muto}, T., {Momose}, M., {et~al.} 2015, \apj, 809, 78

\bibitem[{{Kataoka} {et~al.}(2016{\natexlab{b}}){Kataoka}, {Tsukagoshi},
  {Momose}, {Nagai}, {Muto}, {Dullemond}, {Pohl}, {Fukagawa}, {Shibai},
  {Hanawa}, \& {Murakawa}}]{Kataoka16hd}
{Kataoka}, A., {Tsukagoshi}, T., {Momose}, M., {et~al.} 2016{\natexlab{b}},
  \apjl, 831, L12

\bibitem[{{Kauffmann} {et~al.}(2008){Kauffmann}, {Bertoldi}, {Bourke}, {Evans},
  \& {Lee}}]{Kauffmann08}
{Kauffmann}, J., {Bertoldi}, F., {Bourke}, T.~L., {Evans}, II, N.~J., \& {Lee},
  C.~W. 2008, \aap, 487, 993

\bibitem[{{Kauffmann} {et~al.}(2017){Kauffmann}, {Goldsmith}, {Melnick},
  {Tolls}, {Guzman}, \& {Menten}}]{Kauffmann17}
{Kauffmann}, J., {Goldsmith}, P.~F., {Melnick}, G., {et~al.} 2017, \aap, 605,
  L5

\bibitem[{{Klassen} {et~al.}(2017){Klassen}, {Pudritz}, \& {Kirk}}]{Klassen17}
{Klassen}, M., {Pudritz}, R.~E., \& {Kirk}, H. 2017, \mnras, 465, 2254

\bibitem[{{Koch} \& {Rosolowsky}(2015)}]{KochRosolowsky15}
{Koch}, E.~W., \& {Rosolowsky}, E.~W. 2015, \mnras, 452, 3435

\bibitem[{{Koch} {et~al.}(2018){Koch}, {Tang}, {Ho}, {Yen}, {Su}, \&
  {Takakuwa}}]{Koch18}
{Koch}, P.~M., {Tang}, Y.-W., {Ho}, P.~T.~P., {et~al.} 2018, \apj, 855, 39

\bibitem[{{Kolmogorov}(1933)}]{Kolmogorov33}
{Kolmogorov}, A.~N. 1933, Giornale dell'Istituto Italiano degli Attuari, 4, 83

\bibitem[{{Kuan} {et~al.}(2004){Kuan}, {Huang}, {Charnley}, {Hirano},
  {Takakuwa}, {Wilner}, {Liu}, {Ohashi}, {Bourke}, {Qi}, \& {Zhang}}]{Kuan04}
{Kuan}, Y.-J., {Huang}, H.-C., {Charnley}, S.~B., {et~al.} 2004, \apjl, 616,
  L27

\bibitem[{{Kwon} {et~al.}(2009){Kwon}, {Looney}, {Mundy}, {Chiang}, \&
  {Kemball}}]{Kwon09}
{Kwon}, W., {Looney}, L.~W., {Mundy}, L.~G., {Chiang}, H.-F., \& {Kemball},
  A.~J. 2009, \apj, 696, 841

\bibitem[{{Kwon} {et~al.}(2018){Kwon}, {Stephens}, {Tobin}, {Looney}, {Li},
  {van der Tak}, \& {Crutcher}}]{Kwon18}
{Kwon}, W., {Stephens}, I., {Tobin}, J., {et~al.} 2018, ArXiv e-prints,
  arXiv:1805.07348

\bibitem[{{Lazarian}(2007)}]{Lazarian07}
{Lazarian}, A. 2007, \jqsrt, 106, 225

\bibitem[{{Lazarian} \& {Hoang}(2007)}]{LazarianHoang07}
{Lazarian}, A., \& {Hoang}, T. 2007, \mnras, 378, 910

\bibitem[{{Lee} {et~al.}(2018){Lee}, {Li}, {Ching}, {Lai}, \& {Yang}}]{Lee18}
{Lee}, C.-F., {Li}, Z.-Y., {Ching}, T.-C., {Lai}, S.-P., \& {Yang}, H. 2018,
  \apj, 854, 56

\bibitem[{{Liu} {et~al.}(2018){Liu}, {Hasegawa}, {Ching}, {Lai}, {Hirano}, \&
  {Rao}}]{Liu18}
{Liu}, H.~B., {Hasegawa}, Y., {Ching}, T.-C., {et~al.} 2018, \aap, 617, A3

\bibitem[{{Loinard} {et~al.}(2013){Loinard}, {Zapata}, {Rodr{\'{\i}}guez},
  {Pech}, {Chandler}, {Brogan}, {Wilner}, {Ho}, {Parise}, {Hartmann}, {Zhu},
  {Takahashi}, \& {Trejo}}]{Loinard13}
{Loinard}, L., {Zapata}, L.~A., {Rodr{\'{\i}}guez}, L.~F., {et~al.} 2013,
  \mnras, 430, L10

\bibitem[{{Machida} {et~al.}(2008){Machida}, {Tomisaka}, {Matsumoto}, \&
  {Inutsuka}}]{Machida08}
{Machida}, M.~N., {Tomisaka}, K., {Matsumoto}, T., \& {Inutsuka}, S.-i. 2008,
  \apj, 677, 327

\bibitem[{{Massey}(1951)}]{Massey51}
{Massey}, F.~A. 1951, Journal of the American Statistical Association, 46, 68

\bibitem[{{Masson} {et~al.}(2016){Masson}, {Chabrier}, {Hennebelle}, {Vaytet},
  \& {Commer{\c c}on}}]{Masson16}
{Masson}, J., {Chabrier}, G., {Hennebelle}, P., {Vaytet}, N., \& {Commer{\c
  c}on}, B. 2016, \aap, 587, A32

\bibitem[{{Maury} {et~al.}(2018){Maury}, {Girart}, {Zhang}, {Hennebelle},
  {Keto}, {Rao}, {Lai}, {Ohashi}, \& {Galametz}}]{Maury18}
{Maury}, A.~J., {Girart}, J.~M., {Zhang}, Q., {et~al.} 2018, \mnras, 477, 2760

\bibitem[{{McKee} \& {Ostriker}(2007)}]{McKeeOstriker07}
{McKee}, C.~F., \& {Ostriker}, E.~C. 2007, \araa, 45, 565

\bibitem[{{Mestel} \& {Spitzer}(1956)}]{MestelSpitzer56}
{Mestel}, L., \& {Spitzer}, Jr., L. 1956, \mnras, 116, 503

\bibitem[{{Mestel} \& {Strittmatter}(1967)}]{MestelStrittmatter67}
{Mestel}, L., \& {Strittmatter}, P.~A. 1967, \mnras, 137, 95

\bibitem[{{Mocz} \& {Burkhart}(2018)}]{MoczBurkhart18}
{Mocz}, P., \& {Burkhart}, B. 2018, \mnras, 480, 3916

\bibitem[{{Monsch} {et~al.}(2018){Monsch}, {Pineda}, {Liu}, {Zucker}, {How-Huan
  Chen}, {Pattle}, {Offner}, {Di Francesco}, {Ginsburg}, {Ercolano}, {Arce},
  {Friesen}, {Kirk}, {Caselli}, \& {Goodman}}]{Monsch18}
{Monsch}, K., {Pineda}, J.~E., {Liu}, H.~B., {et~al.} 2018, \apj, 861, 77

\bibitem[{{Mouschovias}(1976)}]{Mouschovias76}
{Mouschovias}, T.~C. 1976, \apj, 206, 753

\bibitem[{{Mundy} {et~al.}(1992){Mundy}, {Wootten}, {Wilking}, {Blake}, \&
  {Sargent}}]{Mundy92}
{Mundy}, L.~G., {Wootten}, A., {Wilking}, B.~A., {Blake}, G.~A., \& {Sargent},
  A.~I. 1992, \apj, 385, 306

\bibitem[{{Nakamura} \& {Li}(2008)}]{NakamuraLi08}
{Nakamura}, F., \& {Li}, Z.-Y. 2008, \apj, 687, 354

\bibitem[{{Natta} {et~al.}(2007){Natta}, {Testi}, {Calvet}, {Henning},
  {Waters}, \& {Wilner}}]{Natta07}
{Natta}, A., {Testi}, L., {Calvet}, N., {et~al.} 2007, Protostars and Planets
  V, 767

\bibitem[{{Offner} {et~al.}(2016){Offner}, {Dunham}, {Lee}, {Arce}, \&
  {Fielding}}]{Offner16}
{Offner}, S.~S.~R., {Dunham}, M.~M., {Lee}, K.~I., {Arce}, H.~G., \&
  {Fielding}, D.~B. 2016, \apjl, 827, L11

\bibitem[{{Ohashi} {et~al.}(2018){Ohashi}, {Kataoka}, {Nagai}, {Momose},
  {Muto}, {Hanawa}, {Fukagawa}, {Tsukagoshi}, {Murakawa}, \&
  {Shibai}}]{Ohashi18}
{Ohashi}, S., {Kataoka}, A., {Nagai}, H., {et~al.} 2018, \apj, 864, 81

\bibitem[{{Ormel} {et~al.}(2011){Ormel}, {Min}, {Tielens}, {Dominik}, \&
  {Paszun}}]{Ormel11}
{Ormel}, C.~W., {Min}, M., {Tielens}, A.~G.~G.~M., {Dominik}, C., \& {Paszun},
  D. 2011, \aap, 532, A43+

\bibitem[{{Ortiz-Le{\'o}n} {et~al.}(2017){Ortiz-Le{\'o}n}, {Loinard},
  {Kounkel}, {Dzib}, {Mioduszewski}, {Rodr{\'{\i}}guez}, {Torres},
  {Gonz{\'a}lez-L{\'o}pezlira}, {Pech}, {Rivera}, {Hartmann}, {Boden}, {Evans},
  {Brice{\~n}o}, {Tobin}, {Galli}, \& {Gudehus}}]{OrtizLeon17}
{Ortiz-Le{\'o}n}, G.~N., {Loinard}, L., {Kounkel}, M.~A., {et~al.} 2017, \apj,
  834, 141

\bibitem[{{Ossenkopf} \& {Henning}(1994)}]{Ossenkopf94}
{Ossenkopf}, V., \& {Henning}, T. 1994, \aap, 291, 943

\bibitem[{{Ostriker} {et~al.}(2001){Ostriker}, {Stone}, \&
  {Gammie}}]{Ostriker01}
{Ostriker}, E.~C., {Stone}, J.~M., \& {Gammie}, C.~F. 2001, \apj, 546, 980

\bibitem[{{Oya} {et~al.}(2016){Oya}, {Sakai}, {L{\'o}pez-Sepulcre}, {Watanabe},
  {Ceccarelli}, {Lefloch}, {Favre}, \& {Yamamoto}}]{Oya16}
{Oya}, Y., {Sakai}, N., {L{\'o}pez-Sepulcre}, A., {et~al.} 2016, \apj, 824, 88

\bibitem[{{Oya} {et~al.}(2018){Oya}, {Moriwaki}, {Onishi}, {Sakai},
  {L{\'o}pez-Sepulcre}, {Favre}, {Watanabe}, {Ceccarelli}, {Lefloch}, \&
  {Yamamoto}}]{Oya18}
{Oya}, Y., {Moriwaki}, K., {Onishi}, S., {et~al.} 2018, \apj, 854, 96

\bibitem[{{Padoan} {et~al.}(2001){Padoan}, {Goodman}, {Draine}, {Juvela},
  {Nordlund}, \& {R{\"o}gnvaldsson}}]{Padoan01}
{Padoan}, P., {Goodman}, A., {Draine}, B.~T., {et~al.} 2001, \apj, 559, 1005

\bibitem[{{Palmeirim} {et~al.}(2013){Palmeirim}, {Andr{\'e}}, {Kirk},
  {Ward-Thompson}, {Arzoumanian}, {K{\"o}nyves}, {Didelon}, {Schneider},
  {Benedettini}, {Bontemps}, {Di Francesco}, {Elia}, {Griffin}, {Hennemann},
  {Hill}, {Martin}, {Men'shchikov}, {Molinari}, {Motte}, {Nguyen Luong},
  {Nutter}, {Peretto}, {Pezzuto}, {Roy}, {Rygl}, {Spinoglio}, \&
  {White}}]{Palmeirim13}
{Palmeirim}, P., {Andr{\'e}}, P., {Kirk}, J., {et~al.} 2013, \aap, 550, A38

\bibitem[{{Panopoulou} {et~al.}(2016){Panopoulou}, {Psaradaki}, \&
  {Tassis}}]{Panopoulou16}
{Panopoulou}, G.~V., {Psaradaki}, I., \& {Tassis}, K. 2016, \mnras, 462, 1517

\bibitem[{{Pattle} {et~al.}(2017){Pattle}, {Ward-Thompson}, {Berry},
  {Hatchell}, {Chen}, {Pon}, {Koch}, {Kwon}, {Kim}, {Bastien}, {Cho},
  {Coud{\'e}}, {Di Francesco}, {Fuller}, {Furuya}, {Graves}, {Johnstone},
  {Kirk}, {Kwon}, {Lee}, {Matthews}, {Mottram}, {Parsons}, {Sadavoy},
  {Shinnaga}, {Soam}, {Hasegawa}, {Lai}, {Qiu}, \& {Friberg}}]{Pattle17}
{Pattle}, K., {Ward-Thompson}, D., {Berry}, D., {et~al.} 2017, \apj, 846, 122

\bibitem[{{Pech} {et~al.}(2010){Pech}, {Loinard}, {Chandler},
  {Rodr{\'{\i}}guez}, {D'Alessio}, {Brogan}, {Wilner}, \& {Ho}}]{Pech10}
{Pech}, G., {Loinard}, L., {Chandler}, C.~J., {et~al.} 2010, \apj, 712, 1403

\bibitem[{{Pineda} {et~al.}(2012){Pineda}, {Maury}, {Fuller}, {Testi},
  {Garc{\'{\i}}a-Appadoo}, {Peck}, {Villard}, {Corder}, {van Kempen}, {Turner},
  {Tachihara}, \& {Dent}}]{Pineda12}
{Pineda}, J.~E., {Maury}, A.~J., {Fuller}, G.~A., {et~al.} 2012, \aap, 544, L7

\bibitem[{{Pinte} {et~al.}(2008){Pinte}, {Padgett}, {M{\'e}nard},
  {Stapelfeldt}, {Schneider}, {Olofsson}, {Pani{\'c}}, {Augereau},
  {Duch{\^e}ne}, {Krist}, {Pontoppidan}, {Perrin}, {Grady}, {Kessler-Silacci},
  {van Dishoeck}, {Lommen}, {Silverstone}, {Hines}, {Wolf}, {Blake}, {Henning},
  \& {Stecklum}}]{Pinte10}
{Pinte}, C., {Padgett}, D.~L., {M{\'e}nard}, F., {et~al.} 2008, \aap, 489, 633

\bibitem[{{Planck Collaboration} {et~al.}(2016){Planck Collaboration}, {Ade},
  {Aghanim}, {Alves}, {Arnaud}, {Arzoumanian}, {Ashdown}, {Aumont},
  {Baccigalupi}, \& et~al.}]{PlanckB15}
{Planck Collaboration}, {Ade}, P.~A.~R., {Aghanim}, N., {et~al.} 2016, \aap,
  586, A138

\bibitem[{{Pohl} {et~al.}(2016){Pohl}, {Kataoka}, {Pinilla}, {Dullemond},
  {Henning}, \& {Birnstiel}}]{Pohl16}
{Pohl}, A., {Kataoka}, A., {Pinilla}, P., {et~al.} 2016, \aap, 593, A12

\bibitem[{{Price} \& {Bate}(2007)}]{PriceBate07}
{Price}, D.~J., \& {Bate}, M.~R. 2007, \mnras, 377, 77

\bibitem[{{Qiu} {et~al.}(2014){Qiu}, {Zhang}, {Menten}, {Liu}, {Tang}, \&
  {Girart}}]{Qiu14}
{Qiu}, K., {Zhang}, Q., {Menten}, K.~M., {et~al.} 2014, \apjl, 794, L18

\bibitem[{{Rao} {et~al.}(2014){Rao}, {Girart}, {Lai}, \& {Marrone}}]{Rao14}
{Rao}, R., {Girart}, J.~M., {Lai}, S.-P., \& {Marrone}, D.~P. 2014, \apjl, 780,
  L6

\bibitem[{{Rao} {et~al.}(2009){Rao}, {Girart}, {Marrone}, {Lai}, \&
  {Schnee}}]{Rao09}
{Rao}, R., {Girart}, J.~M., {Marrone}, D.~P., {Lai}, S.-P., \& {Schnee}, S.
  2009, \apj, 707, 921

\bibitem[{{Ricci} {et~al.}(2010){Ricci}, {Testi}, {Natta}, {Neri}, {Cabrit}, \&
  {Herczeg}}]{Ricci10}
{Ricci}, L., {Testi}, L., {Natta}, A., {et~al.} 2010, \aap, 512, A15

\bibitem[{{Sadavoy} {et~al.}(2018){Sadavoy}, {Myers}, {Stephens}, {Tobin},
  {Commer{\c c}on}, {Henning}, {Looney}, {Kwon}, {Segura-Cox}, \&
  {Harris}}]{Sadavoy18}
{Sadavoy}, S.~I., {Myers}, P.~C., {Stephens}, I.~W., {et~al.} 2018, \apj, 859,
  165

\bibitem[{{Santos} {et~al.}(2016){Santos}, {Busquet}, {Franco}, {Girart}, \&
  {Zhang}}]{Santos16}
{Santos}, F.~P., {Busquet}, G., {Franco}, G.~A.~P., {Girart}, J.~M., \&
  {Zhang}, Q. 2016, \apj, 832, 186

\bibitem[{{Sch{\"o}ier} {et~al.}(2002){Sch{\"o}ier}, {J{\o}rgensen}, {van
  Dishoeck}, \& {Blake}}]{Schoier02}
{Sch{\"o}ier}, F.~L., {J{\o}rgensen}, J.~K., {van Dishoeck}, E.~F., \& {Blake},
  G.~A. 2002, \aap, 390, 1001

\bibitem[{{Seifried} {et~al.}(2013){Seifried}, {Banerjee}, {Pudritz}, \&
  {Klessen}}]{Seifried13}
{Seifried}, D., {Banerjee}, R., {Pudritz}, R.~E., \& {Klessen}, R.~S. 2013,
  \mnras, 432, 3320

\bibitem[{{Seifried} {et~al.}(2015){Seifried}, {Banerjee}, {Pudritz}, \&
  {Klessen}}]{Seifried15}
---. 2015, \mnras, 446, 2776

\bibitem[{{Seifried} {et~al.}(2018){Seifried}, {Walch}, {Reissl}, \&
  {Ib{\'a}{\~n}ez-Mej{\'{\i}}a}}]{Seifried18}
{Seifried}, D., {Walch}, S., {Reissl}, S., \& {Ib{\'a}{\~n}ez-Mej{\'{\i}}a},
  J.~C. 2018, \apj, 855, 81

\bibitem[{{Shirley}(2015)}]{Shirley15}
{Shirley}, Y.~L. 2015, \pasp, 127, 299

\bibitem[{{Simmons} \& {Stewart}(1985)}]{SimmonsStewart85}
{Simmons}, J.~F.~L., \& {Stewart}, B.~G. 1985, \aap, 142, 100

\bibitem[{{Smirnov}(1948)}]{Smirnov48}
{Smirnov}, N. 1948, The Annals of Mathematical Statistics, 19, 279

\bibitem[{{Soler} \& {Hennebelle}(2017)}]{SolerHennebelle17}
{Soler}, J.~D., \& {Hennebelle}, P. 2017, \aap, 607, A2

\bibitem[{{Soler} {et~al.}(2016){Soler}, {Alves}, {Boulanger}, {Bracco},
  {Falgarone}, {Franco}, {Guillet}, {Hennebelle}, {Levrier}, {Martin}, \&
  {Miville-Desch{\^e}nes}}]{Soler16}
{Soler}, J.~D., {Alves}, F., {Boulanger}, F., {et~al.} 2016, \aap, 596, A93

\bibitem[{{Soler} {et~al.}(2017){Soler}, {Ade}, {Angil{\`e}}, {Ashton},
  {Benton}, {Devlin}, {Dober}, {Fissel}, {Fukui}, {Galitzki}, {Gandilo},
  {Hennebelle}, {Klein}, {Li}, {Korotkov}, {Martin}, {Matthews}, {Moncelsi},
  {Netterfield}, {Novak}, {Pascale}, {Poidevin}, {Santos}, {Savini}, {Scott},
  {Shariff}, {Thomas}, {Tucker}, {Tucker}, \& {Ward-Thompson}}]{Soler17}
{Soler}, J.~D., {Ade}, P.~A.~R., {Angil{\`e}}, F.~E., {et~al.} 2017, \aap, 603,
  A64

\bibitem[{{Stephens} {et~al.}(2013){Stephens}, {Looney}, {Kwon}, {Hull},
  {Plambeck}, {Crutcher}, {Chapman}, {Novak}, {Davidson}, {Vaillancourt},
  {Shinnaga}, \& {Matthews}}]{Stephens13}
{Stephens}, I.~W., {Looney}, L.~W., {Kwon}, W., {et~al.} 2013, \apjl, 769, L15

\bibitem[{{Stephens} {et~al.}(2017{\natexlab{a}}){Stephens}, {Dunham}, {Myers},
  {Pokhrel}, {Sadavoy}, {Vorobyov}, {Tobin}, {Pineda}, {Offner}, {Lee},
  {Kristensen}, {J{\o}rgensen}, {Goodman}, {Bourke}, {Arce}, \&
  {Plunkett}}]{Stephens17outflow}
{Stephens}, I.~W., {Dunham}, M.~M., {Myers}, P.~C., {et~al.}
  2017{\natexlab{a}}, \apj, 846, 16

\bibitem[{{Stephens} {et~al.}(2017{\natexlab{b}}){Stephens}, {Yang}, {Li},
  {Looney}, {Kataoka}, {Kwon}, {Fern{\'a}ndez-L{\'o}pez}, {Hull}, {Hughes},
  {Segura-Cox}, {Mundy}, {Crutcher}, \& {Rao}}]{Stephens17}
{Stephens}, I.~W., {Yang}, H., {Li}, Z.-Y., {et~al.} 2017{\natexlab{b}}, \apj,
  851, 55

\bibitem[{Stephens(1974)}]{Stephens74}
Stephens, M.~A. 1974, Journal of American Statistical Association, 69, 730

\bibitem[{{Tazaki} {et~al.}(2017){Tazaki}, {Lazarian}, \& {Nomura}}]{Tazaki17}
{Tazaki}, R., {Lazarian}, A., \& {Nomura}, H. 2017, \apj, 839, 56

\bibitem[{{Testi} {et~al.}(2014){Testi}, {Birnstiel}, {Ricci}, {Andrews},
  {Blum}, {Carpenter}, {Dominik}, {Isella}, {Natta}, {Williams}, \&
  {Wilner}}]{Testi14}
{Testi}, L., {Birnstiel}, T., {Ricci}, L., {et~al.} 2014, Protostars and
  Planets VI, 339


\bibitem[{{Tomisaka}(2011)}]{Tomisaka11}
{Tomisaka}, K. 2011, \pasj, 63, 147

\bibitem[{{Vaillancourt}(2006)}]{Vaillancourt06}
{Vaillancourt}, J.~E. 2006, \pasp, 118, 1340


\bibitem[{{Wootten}(1989)}]{Wootten89}
{Wootten}, A. 1989, \apj, 337, 858

\bibitem[{{Yang} {et~al.}(2016{\natexlab{a}}){Yang}, {Li}, {Looney}, \&
  {Stephens}}]{Yang16}
{Yang}, H., {Li}, Z.-Y., {Looney}, L., \& {Stephens}, I. 2016{\natexlab{a}},
  \mnras, 456, 2794

\bibitem[{{Yang} {et~al.}(2016{\natexlab{b}}){Yang}, {Li}, {Looney}, {Cox},
  {Tobin}, {Stephens}, {Segura-Cox}, \& {Harris}}]{Yang16_iras4a}
{Yang}, H., {Li}, Z.-Y., {Looney}, L.~W., {et~al.} 2016{\natexlab{b}}, \mnras,
  460, 4109

\bibitem[{{Yang} {et~al.}(2017){Yang}, {Li}, {Looney}, {Girart}, \&
  {Stephens}}]{Yang17}
{Yang}, H., {Li}, Z.-Y., {Looney}, L.~W., {Girart}, J.~M., \& {Stephens}, I.~W.
  2017, \mnras, 472, 373

\bibitem[{{Zapata} {et~al.}(2013){Zapata}, {Loinard}, {Rodr{\'{\i}}guez},
  {Hern{\'a}ndez-Hern{\'a}ndez}, {Takahashi}, {Trejo}, \& {Parise}}]{Zapata13}
{Zapata}, L.~A., {Loinard}, L., {Rodr{\'{\i}}guez}, L.~F., {et~al.} 2013,
  \apjl, 764, L14

\end{thebibliography}

\end{document}